\documentclass[
preprint,
nofootinbib,
amsmath,amssymb,
 aps,
]{revtex4-1}

\usepackage{graphicx}
\usepackage{dcolumn}
\usepackage{bm}
\usepackage{array} 
\usepackage[pdfpagelabels,plainpages=false,bookmarks=true,colorlinks,linkcolor=red,urlcolor=blue,citecolor=blue]{hyperref}

\usepackage{float}
\usepackage{cleveref}

\usepackage{amsthm}

\def\be{\begin{equation}}
\def\ee{\end{equation}}
\def\ba{\begin{eqnarray}}
\def\ea{\end{eqnarray}}
\def\sdg{Schr\"odinger~}

\newcommand{\ii}{\mathrm{i}}

\newcommand{\ket}[1]{\lvert #1 \rangle}                  
\newcommand{\bra}[1]{\langle #1 \rvert}
\newcommand{\inner}[2]{\langle #1|#2\rangle}
\newcommand{\abs}[1]{\lvert #1 \rvert}
\newcommand{\norm}[1]{\lVert #1 \rVert}

\DeclareMathOperator{\Tr}{Tr}

\usepackage{tikz,xcolor,graphicx}
\usepackage{CJKutf8}
\begin{document}

\title{Variational method for $\mathbb{Z}_K$ wavefunctions in spin-$J$ PXP model}
\begin{CJK*}{UTF8}{gbsn}
\author{Zhigang Hu(胡志刚)}
\affiliation{International Center for Quantum Materials, School of Physics, Peking University,  Beijing 100871, China}
\author{Biao Wu(吴飙)}
\email{wubiao@pku.edu.cn}
\affiliation{International Center for Quantum Materials, School of Physics, Peking University,  Beijing 100871, China}
\affiliation{Wilczek Quantum Center, Shanghai Institute for Advanced Studies, 
Shanghai 201315, China}

\date{\today}

\begin{abstract}
We investigate the approach of time-dependent variational principle (TDVP) for 
the one-dimensional spin-$J$ PXP model with detuning, which is relevant for programmable Rydberg atom arrays.  
The variational manifold is chosen as the minimally entangled $\mathbb{Z}_K$  matrix-product-states (MPS). 
We demonstrate that variational dynamics and variational error can be expressed 
as rapidly convergent series in the thermodynamic limit. 
In particular, for $J=1/2$ and the limiting case $J\rightarrow \infty$, 
the TDVP results become exact and  significantly simplified. 
\end{abstract}

\maketitle
\end{CJK*}
\section{Introduction}
Recently, advancements in controllable synthetic quantum simulator have opened new opportunities for experimental investigation of non-equilibrium quantum dynamics. Rydberg-atom array trapped in optical tweezers, with their flexible spatial arrangement, presents a promising platform to simulate quantum many-body system in a programmable way. These Rydberg-atom array systems have already demonstrated  potentials in the exploration of exotic phase of matter \cite{Lukin2021-spinliquid}, quantum optimization algorithm \cite{pichler2018} and quantum optimal control \cite{Serbyn2022}. Additionally, they have been used for the study of quantum phase transition with competing crystalline orders \cite{Sachdev2002,Lukin2017,Subir2020} and investigate the quantum non-equilibrium phenomena such as intriguing quantum many-body scarring \cite{Lukin2017,WWHo2019,Turner2018,Serbyn2021}. A key feature of these experiments is the Rydberg blockade, which prevents simultaneous excitations of neighboring atoms. The effective PXP Hamiltonian, which captures the blockade mechanism, is non-integrable \cite{Sachdev2002,WWHo2019,Khemani2019}, making it challenging to obtain an analytical understanding of the experimental results.

The variational method offers a feasible route to describe the Rydberg physics.  The idea of using variational approach to quantum many body dynamics dates back to Dirac \cite{Dirac_1930}. In this method, the exact evolution of the variational state is projected onto the tangent plane of the variational manifold. The variational parameters are then updated along this projection direction.  
The procedure is now known as time-dependent variational principle (TDVP) \cite{Haegeman2011,Hackl2020,Verstraete2019}. The equations of motion obtained can be regarded as an effective semiclassical description of quantum dynamics, as long as the variational error is not significant. 

The TDVP method has successfully explained long-time many-body oscillations in 1D Rydberg atom array. 
The variational ansatz used is the matrix product state (MPS) with $\mathbb{Z}_2$ discrete translational symmetry, subject to additional constraint that avoids simultaneous neighbour excitation\textemdash akin to the Gutzwiller projection. The resulting analytical tractable equations of motion attribute the revival to unstable, short-period periodic orbits in phase space \cite{WWHo2019,Serbyn2020}, reminiscent of single-particle quantum scars in stadium billiards \cite{Heller1976}. However, the phase transition and non-equilibrium dynamics observed in Rydberg atom array experiments present rich sublattice symmetries beyond simple $\mathbb{Z}_2$ ansatz \cite{Lukin2017}.   To extend the approach to system with more general $\mathbb{Z}_K$ symmetry, 
a generalized MPS ansatz with $\mathbb{Z}_K$ periodicity is required 
for TDVP method. This generalization is nontrivial, as the standard transfer matrix technique becomes much more complex for large periodicities.

In this article, we study the spin-$J$ PXP model, which generalizes the spin $1/2$ PXP model and includes it as a special case. Higher-spin models  have long been fertile ground for exploring exotic physics \cite{AKLT1987} and their realization has attracted significant experimental attention recently \cite{ZhangY2024,Norman2024}. 
Employing the TDVP method, we investigate the model by adopting the minimally-entangled MPS ansatz (with nontrivial bond dimension $D=2$), which allows for tensor contractions to be performed even by hand. The variational MPS with $\mathbb{Z}_K$ discrete translational symmetry  inherently encodes the Rydberg blockade constraint. 
With mathematical induction, we explicitly solve  the transfer matrix for a $\mathbb{Z}_K$ unit cell. The dominant eigenvectors of the transfer matrix satisfy reduction formulae  when the eigenvectors are contracted with the transfer matrix over arbitrary interval of lattice sites. These observations facilitate analytical calculations of variational dynamics, and combined with the energy variance,  allow for the estimation of variational error rate, although more complex calculations involved than those of variational dynamics.
Notably, both the variational dynamics and error rate are expressed as rapidly convergent series, enabling efficient truncation.

The article is organized as follows. In Sec. \ref{sec:tdvp-general}, we introduce the TDVP method for general variational ansatz, and in Sec. \ref{sec:spinJHam}, we describe the spin-$J$ PXP model. In Sec. \ref{sec:transfer-matrix-formalism}, we specify the $\mathbb{Z}_K$ variational MPS ansatz and derive the transfer matrix over subsequent lattice sites. The reduction formulae are also proposed such that the contraction between the dominant eigenvectors and the transfer matrix becomes straightforward. In Sec. \ref{sec:energy}, we present an analytical expression for the variational energy. In Sec. \ref{sec:tdvp-results}, we work out the variational dynamics and variational error rate in the form of rapidly convergent series. We show that TDVP results are exact and compact for spin $\frac{1}{2}$ and large spin limit. In Sec. \ref{sec:discussion}, we discuss several open questions and future directions for this research.

\section{Time-dependent variational principle}\label{sec:tdvp-general}
The TDVP equations of motion are determined by taking the extreme of the action $S=\int\mathcal{L}dt=\int \mathrm{i}(\langle\psi|\dot{\psi}\rangle-\langle\dot{\psi}|\psi\rangle)/2-\langle\psi|H|\psi\rangle dt$. The action is invariant under the gauge transformation $\ket{\psi(t)}\rightarrow e^{\ii \alpha(t)}\ket{\psi(t)}$ with a phase factor $e^{\ii \alpha(t)}$. We fix the redundant degree of freedom such that the ``mean'' \sdg equation $\bra{\psi}(\frac{d}{dt}+\ii H )\ket{\psi}$ holds. It has been already noticed that this gauge choice removes all disconnected correlations in TDVP equations and quantum leakage \cite{Serbyn2020}.

Let the variational parameters be $z_i$, the equations of motion of $z_i$ read
\begin{equation}
    \sum_{j} \dot{\mu}_j\text{Im}G_{ij}^{c} = - \text{Re}\bra{\partial_{\mu_i}\psi}H\ket{\psi}_c\,,
    \label{eq:eom}
\end{equation}
where $G_{ij}^{c}\equiv \inner{\partial_{\mu_i}\psi}{\partial_{\mu_j}\psi}-  \inner{\partial_{\mu_i}\psi}{\psi}\inner{\psi}{\partial_{\mu_j}\psi}$ is the connected part of the Gram matrix $G_{ij}=\inner{\partial_{\mu_i}\psi}{\partial_{\mu_j}\psi}$. Similarly, $\bra{\partial_{\mu_i}\psi}H\ket{\psi}_c\equiv \bra{\partial_{\mu_i}\psi}H\ket{\psi}-\inner{\partial_{\mu_i}\psi}{\psi} \bra{\psi} H\ket{\psi}$ has its disconnected part removed. The semiclassical trajectory can be obtained by integrating the TDVP equations Eq.~(\ref{eq:eom}). Along the trajectory, the instantaneous variational error rate, also dubbed as quantum leakage in literature, becomes
\begin{align}
    & \Lambda^2 = \frac{1}{L}\norm{\ket{\dot{\psi}}+\ii H \ket{\psi}}^2 \nonumber\\
    &=\frac{1}{L} \Big\{ \bra{\psi}H^2\ket{\psi}_c - 2 \sum_i \dot{\mu}_i \text{Im}\bra{\partial_{\mu_i}\psi}H\ket{\psi}_c + \sum_{ij} \dot{\mu}_i\dot{\mu}_j \text{Re}G_{ij}^{c}\Big\}\,,
    \label{eq:Qleak}
\end{align}
where $\langle\psi|H^{2}|\psi\rangle_{c}\equiv\langle\psi|H^{2}|\psi\rangle-\langle\psi|H|\psi\rangle^{2}$ is the connected part of the energy variance.  It is noted that we normalize $\Lambda^2 $ with the system size $L$ such that $\Lambda^2 $ is well-defined in thermodynamic limit. Quantum leakage is found to be very useful diagnosis for quantum-classical correspondence. When $\int_{0}^{\tau} \Lambda dt \lesssim 1$, it is reasonable to infer the quantum dynamic can be well described by the semiclassical trajectory within the timescale $\tau$ \cite{evrard2024}.

\section{Spin-$J$ PXP model}\label{sec:spinJHam}
Consider a 1D Rydberg atom array. Each Rydberg atom has internal degree of freedom which we may call it spin.  Let the unexcited state be ``spin-down'' $\ket{0}\equiv \ket{J,-J}$, which is the lowest-lying eigenstate of the spin operator $S^z$. Under the experimental setting, the van der Waals interaction of neighboring atoms is tuned to dominate over the Rabi frequency $\Omega$ and the detuning $\Delta$. In that case, simultaneous excitation of adjacent atoms is energetically unfavorable. Inspired by the experiments, we thus focus on a model Hamiltonian that demonstrate such constrained spin dynamics 
\begin{equation}
    H = \sum_i \Omega_i\mathcal{P} s_i^x \mathcal{P}+\Delta_i s_i^z\,,
    \label{eq:spinJ-pxp}
\end{equation}
where $s_i^x=J^{-1}S_i^x$ and $s_i^z=J^{-1}S_i^z$ are normalized spin-$J$ operators at site $i$. The normalization of spin operator is used such that local energy density is always finite whatever value $J$ takes.  The projector $\mathcal{P}=\prod_i P_{i,i+1} = \prod_i I_i\otimes I_{i+1}-(I_i-P_i)\otimes(I_{i+1}-P_{i+1})$ projects out states with nearest neighboring excitation, where $P_i=|0\rangle_i\langle0|_i$ is projector of unexcited state at site $i$. 
We will restrict our discussion in a constrained Hilbert space $\mathcal{H}$ \cite{Zlatko2018}, spanned by basis states that avoid simultaneous neighbor excitation.

When $J=1/2$, the Hamiltonian (\ref{eq:spinJ-pxp}) reduces to the detuned PXP model in \cite{Pan2023}.  The spin-$J$ model has a particle-hole symmetry $U = e^{\mathrm{i}\pi J\sum_i s_i^{z}}$ when there is no detuning, shown by $U^{-1}s_i^{x}U=-s_i^{x}$. For general parameter setting there is no obvious symmetry. It is known that the spin-$J$ PXP model can exhibit non-ergodic quantum dynamics when $J$ is small \cite{WWHo2019}.

\section{transfer matrix formalism}\label{sec:transfer-matrix-formalism}
We study the quantum dynamics of the spin-$J$ PXP model through TDVP method. 
Following \cite{WWHo2019}, our trial wave function is a MPS 
\begin{equation}
|\Psi\rangle=\text{Tr}\bigg(\prod_{i=1}^{L}A_{i}\bigg)\,,
\end{equation}
with onsite matrix
\begin{equation}
\quad A_{i}(\theta_{i},\phi_{i})=\left(\begin{array}{cc}
P_{i}|\theta_{i},\phi_{i}\rangle & (I_{i}-P_{i})|\theta_{i},\phi_{i}\rangle\\
|0\rangle_{i} & 0
\end{array}\right)\,,
\label{eq:vansatz}
\end{equation}
where $|\theta_{i},\phi_{i}\rangle=e^{\xi_i S_i^{+}-\xi_i^{*}S_i^{-}}|0\rangle_i$ is the spin coherent state with $\xi_i=\frac{\theta_i}{2}e^{-\mathrm{i}\phi_i}$ and $S^{\pm}_i=S^x_i\pm \mathrm{i}S^y_i$. We summarize some important properties of the spin coherent state  in Appendix~\ref{sec:spin coherent state}. For $J=\frac{1}{2}$, the onsite matrix is $A_{i}=\left(\begin{array}{cc}
\cos\frac{\theta_i}{2}|0\rangle_i & e^{-\mathrm{i}\phi_i}\sin\frac{\theta_i}{2}|1\rangle_i \\
|0\rangle_{i} & 0
\end{array}\right)$. Here, $L$ is the system size and periodic boundary condition is used.
The variational wavefunction automatically excludes nearest neighbour excitation as $\mathcal{P} \ket{\Psi}= \ket{\Psi}$, which is revealed by $P_{i,i+1} A_i A_{i+1} = A_i A_{i+1}$, since there is no component $(I_i-P_i)\ket{\theta_i,\phi_i}\otimes (I_{i+1}-P_{i+1})\ket{\theta_{i+1},\phi_{i+1}}$ in matrix elements of $A_i A_{i+1}$.

We will use transfer matrix method to solve the correlation function in equations of motion and quantum leakage. The transfer matrix of a block of adjacent sites from $i$ to $j$ is  $T_{[i,j]}\equiv T_{i}\cdot T_{i+1}\cdots T_{j}$, where $T_i$ is the transfer matrix at site $i$, in terms of $x_i \equiv \cos^{2J}\frac{\theta_i}{2}$
\begin{equation}
    T_i \equiv \sum_{\sigma_i} (A^{\sigma_i}_i)^{*}\otimes A^{\sigma_i}_i = \left(\begin{array}{cccc}
x_i^{2} & 0 & 0 & 1-x_i^{2}\\
x_i & 0 & 0 & 0\\
x_i & 0 & 0 & 0\\
1 & 0 & 0 & 0
\end{array}\right).
\end{equation}
where $\sigma_i$ sums from $-J$ to $J$, $A^{\sigma_i}_i$ is the $\ket{J,\sigma_i}_i$ component of the on-site matrix $A_i$ and $(A^{\sigma_i}_i)^{*}$ is the complex conjugate of $A^{\sigma_i}_i$. Importantly, $T_i$ is quite sparse. After a recursion procedure in Appendix \ref{sec:transfer matrix}, we find that $T_{[i,j]}$ has following compact form
\begin{align}
    T_{[i,j]}=\left(\begin{array}{cccc}
1+\alpha_{[i,j]}+\beta_{[i,j]} & 0 & 0 & -\alpha_{[i,j]}-\beta_{[i,j]}\\
x_{i} (1+\alpha_{[i,j]}) & 0 & 0 & -x_{i}\alpha_{[i,j]} \\
x_{i} (1+\alpha_{[i,j]}) & 0 & 0 & -x_{i}\alpha_{[i,j]} \\
1+\alpha_{[i,j]} & 0 & 0 & -\alpha_{[i,j]}
\end{array}\right)
\label{eq:Tij}
\end{align}
with $\alpha_{[i,j]} = \sum_{m=i}^{j}\prod_{k=m}^{j} (-1+x_k^2)-\prod_{k=i}^{j}(-1+x_k^2)$ and $\beta_{[i,j]} = \prod_{k=i}^{j}(-1+x_k^2)$. The eigenvalues of $T_{[i,j]}$ are $\lambda_1 = 1$, $\lambda_2 = \beta_{[i,j]}$ and $\lambda_3=\lambda_4 = 0$.  The dominating eigenvalue is $\lambda_1=1$, since $\abs{\lambda_2}<1$ unless all $\theta_i=\pi$. Therefore the trial wave function $\ket{\Psi}$ is normalized in the thermodynamic limit. 

We focus on the case where $\ket{\Psi}$ has $\mathbb{Z}_K$ discrete translational symmetry, i.e. $A_{i+K}=A_{i}$.  The variational parameters are $\theta_{i}$ and $\phi_{i}$ for $i=1,\cdots,K$. The phase space is $2K$ dimensional, which admits chaos for $K\geq 2$. Without loss of generality, hereafter we assume $K$ is sufficient large such that the eigenvalue $\lambda_2 = \beta_{[1,K]}$ is negligible. A translational invariant unit is $T_{[i,i+K-1]}$, whose dominant right and left eigenvectors are
\begin{equation}
    |r_i) =(1,x_i,x_i,1)^{\text{T}}, \quad (l_i|=(\eta_i,0,0,1-\eta_i )\,,
\end{equation}
where 
\begin{equation}
    \eta_i = 1+\sum_{m\leq i-1}\prod_{j= m}^{i-1}(-1+x_j^2)\,.
    \label{eq:eta-def}
\end{equation}
We can enforce the periodicity of $x_i$ in the large unit cell to recover the small $K$ case, 
which can be justified from the general formula for arbitrary periodicity Eq.~(\ref{eq:etai}).
We list small $K$ results explicitly
\begin{equation}
    \eta_i = \begin{cases}
        \frac{1}{2-\cos^{4J}\frac{\theta_i}{2}} & K=1\\
        \frac{\cos^{4J}\frac{\theta_{i-1}}{2}}{1-(-1+\cos^{4J}\frac{\theta_{i-1}}{2})(-1+\cos^{4J}\frac{\theta_{i}}{2} )} & K=2\\
        \frac{1+(-1+\cos^{4J}\frac{\theta_{i-1}}{2})\cos^{4J}\frac{\theta_{i-2}}{2}}{1-(-1+\cos^{4J}\frac{\theta_{i-1}}{2})(-1+\cos^{4J}\frac{\theta_{i-2}}{2})(-1+\cos^{4J}\frac{\theta_{i}}{2})} & K=3
    \end{cases}
    \label{eq:eta-small-K}
\end{equation}
The contraction between dominant eigenvectors and the transfer matrix $T_{[i,j]}$ is given by reduction formulae (see Eq.~(\ref{eq:reduction1}) and (\ref{eq:reduction2}))
\begin{equation}
    (l_i|T_{[i,j]}  = (l_{j+1}|,\quad T_{[i,j]}|r_{j+1})=|r_i)\,.
    \label{eq:reduction-formula}
\end{equation}
The reduction formulae simply represent different choice of the $\mathbb{Z}_K$ unit cell in the Perron-Frobenius theorem, which tells us $\lim_{m\rightarrow\infty}(T_{[i,i+K-1]})^{m} = |r_i)(l_i|$ when $T_{[i,i+K-1]}$ has a single dominant eigenvalue $\lambda=1$ with eigenvectors $(l_i|$ and $|r_i)$. When we calculate the expectation value (and correlation functions), we can apply the Perron-Frobenius theorem to both the left and right semi-infinite transfer matrices. By taking trace operation and noticing $(l_i|r_j)=1$ even for $i\neq j$, the validity of the reduction formula can be justified.

\section{variational energy}\label{sec:energy}
When we evaluate the variational energy $\bra{\Psi}H\ket{\Psi}$, the projector $\mathcal{P}$ in $\mathcal{P} s_i^x \mathcal{P}$ in the Hamiltonian can be omitted, as implied by $\mathcal{P}\ket{\Psi} = \ket{\Psi}$. Then we only need to deal with the expectation value of on-site operator. Define the transfer matrix with onsite operator $q_i$ as
\begin{equation}
    T_{q_i} \equiv  \sum_{\sigma,\sigma'} (A^{\sigma}_i)^{*}\otimes q_i^{\sigma,\sigma'} A^{\sigma'}_i \,,
\end{equation}
where $q_i^{\sigma,\sigma'}$ is the matrix element of $q_i$ in the Dicke basis. The explicit matrix elements of $T_{q_i}$ are given in Eq.~(\ref{eq:tran-mat-op}).

Choosing a specific unit cell from $j$ to $j+K-1$ and applying Perron-Frobenius theorem,  we obtain
\begin{equation}
    \bra{\Psi}q_i\ket{\Psi} = (l_j|T_{j}\cdot T_{j+1}\cdots T_{i-1}\cdot T_{q_i} \cdot T_{i+1} \cdots T_{j+K-1} |r_j)\,,
\end{equation}
which, according to the reduction formula Eq.~(\ref{eq:reduction-formula}), can be written as
\begin{align}
\bra{\Psi}q_i\ket{\Psi}
=(l_i|T_{q_i}|r_{i+1}) = \bra{0}_i q_i \ket{0}_i + \eta_i \tilde{h}_{q_i}\,, 
\label{P8eq:onesite-ave}
\end{align}
where we use the shorthand notation
\begin{align}
    \tilde{h}_{q_i}& \equiv  x_i (-1+x_{i+1}) \cdot \Big( \bra{0}_i q_i(1-P_i)\ket{\theta,\phi}_i +  \nonumber\\
    & \bra{\theta,\phi}_i (1-P_i) q_i\ket{0}_i \Big) +  \bra{\theta,\phi}_i q_i\ket{\theta,\phi}_i -\bra{0}_i q_i\ket{0}_i\,.
    \label{eq:htilde}
\end{align}
The onsite operators of the Hamiltonian are simply $\Omega_i s^{x}_i$ and $\Delta_i s^{z}_i$. From Table~\ref{tab:one-site op}, we can calculate the variational energy
\begin{align}
    \bra{\Psi}H\ket{\Psi} & = \sum_{i} -\Delta_i + \eta_i\Big( \Delta_i (1-\cos\theta_i) \nonumber\\
    &+ \Omega_i\sin\theta_i\cos\phi_i ( 1+\cos^{4J-2}\frac{\theta_{i}}{2}(-1+\cos^{2J}\frac{\theta_{i+1}}{2})) \Big)\,.
\end{align}

\section{variational dynamics and error rate} \label{sec:tdvp-results}
While the variational energy reduces to the expectation value of single onsite operators, the TDVP equations of motion and quantum leakage generally involve correlation between onsite operators. However, this is not a major obstacle, as a closed-form expression for the cumulative product of transfer matrices exists (Eq.~(\ref{eq:Tij})).

Let us first compute the variational dynamics. From Eq.~(\ref{eq:eom}), we need to invert the Gram matrix and calculate the partial derivative $\text{Re}\bra{\partial_{\mu_i}\psi}H\ket{\psi}_c$.
A typical term in $G_{\mu_i,\nu_j}^c$ is
\begin{equation}
    (l_i| \bar{\partial}_{\mu}T_i \cdot T_{[i+1,j-1]}\cdot \partial_{\nu}T_{j} |r_{j+1})_c 
\end{equation}
where the subscript `c' denotes removal of disconnected parts $(l_i| \bar{\partial}_{\mu}T_i |r_{i+1}) \cdot (l_j| \partial_{\nu}T_{j} |r_{j+1})$. Here $\bar{\partial}_{\mu}T_i \equiv  \sum_{\sigma} (\partial_{\mu_i} A^{\sigma}_i)^{*}\otimes A^{\sigma}_i$ and $\partial_{\nu}T_{j}\equiv \sum_{\sigma} (A^{\sigma}_j)^{*}\otimes \partial_{\nu_j}A^{\sigma}_j$ represent partial derivatives of transfer matrices.
For details, see Appendix \ref{sec:gram}.  Given that $G_{\theta_i,\theta_j}^c$ and $G_{\phi_i,\phi_j}^c$ are real symmetric, $\text{Im}G^{c}$ is nonzero only off-diagonally, so we focus on $\text{Im}G^{c}_{\theta_i,\phi_j}$.  The compact formula of $(\text{Im}G^{c}_{\bm{\theta},\bm{\phi}})^{-1}$ can be found in Appendix \ref{sec:inv-gram}. 
It is worth noting that $\mathcal{P} \partial_{\mu_i}\ket{\Psi}=\partial_{\mu_i}\ket{\Psi}$ as  there is no component $(I_i-P_i)\ket{\theta_i,\phi_i}\otimes (I_{i+1}-P_{i+1})\ket{\theta_{i+1},\phi_{i+1}}$ in matrix elements of $\partial_{\mu_i}A_i A_{i+1}$ and $A_i \partial_{\mu_{i+1}}A_{i+1}$.  Then we can remove the projector $\mathcal{P}$ in $\text{Re}\bra{\partial_{z_i}\psi}H\ket{\psi}_c$, which essentially reduces the Hamiltonian to the sum of single onsite operator $\Omega_i s_i^x + \Delta_i s_i^z$. A typical term in $\bra{\partial_{\mu_i}\psi}H\ket{\psi}_c$ is
\begin{equation}
    (l_i| (\partial_{\mu}T_i) \cdot T_{[i+1,j-1]}\cdot T_{q_{j}} |r_{j+1})_c \,,
\end{equation}
where the subscript `c' denotes removal of disconnected parts $(l_i| (\partial_{\mu}T_i) |r_{i+1}) \cdot (l_j| T_{q_{j}} |r_{j+1})$. The full calculation is rather lengthy (see Appendix \ref{sec:muH}). Collecting these results, we give explicit results of the TDVP equations of motion in Appendix \ref{sec:tdvp-eom}, which are
\begin{align}
    \dot{\theta_i} &= \frac{\Omega_i \tan\phi_i}{J \sin\theta_i} \tilde{h}_{ s_i^x} + \frac{\eta_{i-1}\Omega_{i-1} \cos^{4J-2}\frac{\theta_{i-1}}{2}\tan\phi_{i-1}\tan\frac{\theta_i}{2}}{\eta_i} \tilde{h}_{ s_{i-1}^x} \nonumber\\
    & + \sum_{j\leq i-2} \frac{\eta_j \Omega_j \cos^{4J-2}\frac{\theta_j}{2}\tan\phi_j\tan\frac{\theta_i}{2}}{\eta_i} \tilde{h}_{ s_j^x}\prod_{m=j+1}^{i-1}\tilde{c}_m
    \label{eq:thetadot-maintext}
\end{align}
\begin{align}
    \dot{\phi_i} &= \frac{\Omega_i \tilde{h}_{s_i^x}+\Delta_i (1-\cos\theta_i)}{J(1-\cos\theta_i)} -\frac{2\cos^{4J-2}\frac{\theta_{i}}{2} \tan\frac{\theta_{i+1}}{2} }{\eta_{i+1}} \tilde{R}_{\theta_{i+1}}\,,\nonumber\\
    & - \frac{2}{J\eta_i\sin\theta_i}\tilde{R}_{\theta_i} - \sum_{j\geq i+2} \frac{2\cos^{4J-2}\frac{\theta_{i}}{2} \tan\frac{\theta_j}{2} }{\eta_{j}} \tilde{R}_{\theta_{j}} \prod_{m=i+1}^{j-1} \tilde{c}_m\,,
    \label{eq:phidot-maintext}
\end{align}
where $\tilde{c}_i \equiv -1+(2J\tan^2\frac{\theta_j}{2}+1)\cos^{4J}\frac{\theta_j}{2}$ and according to Eq.~(\ref{eq:htilde}) $\tilde{h}_{s_i^x}=   \sin\theta_i\cos\phi_i ( 1+\cos^{4J-2}\frac{\theta_{i}}{2}(-1+\cos^{2J}\frac{\theta_{i+1}}{2}))$. Here $\tilde{R}_{\theta_i} \equiv J\Omega_{i-1}\eta_{i-1}\cos^{2J}\frac{\theta_i}{2}\tan\frac{\theta_{i}}{2}\cos^{4J}\frac{\theta_{i-1}}{2}\tan\frac{\theta_{i-1}}{2}\cos\phi_{i-1} +\frac{\eta_i\Omega_i}{2\sin\theta_i} \Big(  \tilde{h}_{s_i^x}  + 2\cos\phi_i(2J-1)(1-\cos\theta_i)\cos^{4J}\frac{\theta_i}{2} (-1+\cos^{2J}\frac{\theta_{i+1}}{2})\tan\frac{\theta_i}{2}\Big)$. Since $\abs{\tilde{c}_i}<1$ it is often reasonable to truncate the summation until $\prod_{m}\tilde{c}_m$ becomes negligible. It is interesting to note that $\tilde{c}_i =0$ for $J=\frac{1}{2}$. In spin-$\frac{1}{2}$ case, the summation term simply vanishes.

It is direct to verify that the variational dynamics is invariant under $\theta_i \rightarrow \theta_i + 2\pi$ for integer $J$ and $\theta_i \rightarrow \theta_i + 4\pi$ for half-integer $J$.  $J=\frac{1}{2}$ is special, as a joint transformation $\theta_i \rightarrow \theta_i + 2\pi,\, \phi_i \rightarrow \phi_i + \pi$ leaves the variational dynamics unchanged.

Next we will evaluate the variational error rate. Given the  Gram matrix $G_{\mu_i,\nu_j}^c$, $\bra{\partial_{\mu_i}\psi}H\ket{\psi}_c$ and the TDVP equations of motion, we can compute the error rate from Eq.~(\ref{eq:Qleak}) once we solve the energy variance $\bra{\Psi}H^2\ket{\Psi}_c$. Unlike in $\bra{\partial_{\mu_i}\psi}H\ket{\psi}_c$, the projector $\mathcal{P}$ can not be simply omitted in the energy variance. It is useful to decompose $\mathcal{P}$ into product of two-site operators $P_{i,i+1}$, which satisfy local relation $P_{i,i+1}A_i\cdot A_{i+1}=A_i \cdot A_{i+1}$ and $A_i^{*}\cdot A_{i+1}^{*}P_{i,i+1}=A_i^{*}\cdot A_{i+1}^{*}$, and commute with $s_j^z$. This allows us to reduce terms involving $s_i^z$ in $\bra{\Psi}H^2\ket{\Psi}_c$ to the correlation of two onsite operators. For instance, $\bra{\Psi} \mathcal{P} s_i^x \mathcal{P}  s_{j}^z \ket{\Psi} = \bra{\Psi}  s_i^x   s_{j}^z \ket{\Psi}$, which follows from first exchanging $P_{i,i+1}$ with $s_{j}^z$ and then acting $P_{i,i+1}$ on $A_i\cdot A_{i+1}$ in the $\ket{\Psi}$ or $A_i^{*}\cdot A_{i+1}^{*}$ in the $\bra{\Psi}$.  A typical term in the correlation of two onsite operators is
\begin{align}
    (l_i|T_{q_i}\cdot T_{[i+1,j-1]}\cdot T_{q_j}|r_{j+1})_c \,,
\end{align}
where the subscript `c' denotes removal of disconnected parts $(l_i|T_{q_i}|r_{i+1})\cdot(l_j|T_{q_j}|r_{j+1})$. For terms like $\bra{\Psi} \mathcal{P} s_i^x \mathcal{P}  \mathcal{P} s_{j}^x \mathcal{P} \ket{\Psi}$, 
we must treat cases differently based on $\abs{j-i}$.  When $\abs{j-i}>1$, 
we can freely apply $P_{i,i+1}$ to at least one of $\ket{\Psi}$ and $\bra{\Psi}$, reducing the case to a correlation of two onsite operators.  When $\abs{j-i}\leq 1$, there is obstruction to apply $P_{i,i+1}$ to either $\ket{\Psi}$ or $\bra{\Psi}$. Taking $j=i+1$ as an example, by direct calculation
\begin{align}
    &\bra{\Psi} \mathcal{P} s_i^x \mathcal{P} \mathcal{P} s_{i+1}^x \mathcal{P}  \ket{\Psi} = \bra{\Psi} \mathcal{P} s_i^x \mathcal{P} s_{i+1}^x   \ket{\Psi} \nonumber\\
    & = \bra{\Psi} P_{i-1,i}P_{i,i+1} s^x_i  P_{i-1,i}P_{i,i+1}  s_{i+1}^x   \ket{\Psi}\,.
\end{align}
This simplifies to
\begin{equation}
    \bra{\Psi} P_{i-1} s^x_i  P_{i+1}  s_{i+1}^x   \ket{\Psi}\,,
    \label{eq:xxterm}
\end{equation}
since $P_{i}  s^x_i  P_{i}=0$.
Then we see that the onsite operator $s_i^x$ is ``dressed'' by $P_{i-1}$ and $P_{i+1}$. 
A typical term in Eq.~(\ref{eq:xxterm}) is
\begin{equation}
    (l_{i-1}|T_{P_{i-1}}\cdot T_{s_i^x}\cdot T_{P_{i+1}s^x_{i+1}}|r_{i+2})_c\,,
\end{equation}
where the subscript `c' denotes the removal of the disconnected part $(l_i|T_{s_i^x}|r_{i+1})\cdot  (l_{i+1}|T_{s^x_{i+1}}|r_{i+2})$. For further details, the full calculation of the quantum leakage is in Appendix \ref{sec:qleak}.


In previous discussion, we outlined the procedure for solving TDVP equations of motion and quantum leakage in the spin-$J$ PXP model. It is natural to explore two extreme limits: $J=\frac{1}{2}$ and $J\rightarrow \infty$. These limits are significant because they represent fundamentally different physical regimes. For $J=\frac{1}{2}$, the system is a set of interacting qubits with strong quantum fluctuation. On the other hand, in the large spin limit, the system enters a highly classical-like regime \cite{Lieb1973}. Note that the large spin limit is taken after the thermodynamic limit $L\rightarrow \infty$. 

When $J=\frac{1}{2}$, the variational dynamics is surprisingly simple as  $\tilde{c}_i =0$
\begin{equation}
    \dot{\theta}_i =2 \Omega_i \sin\phi_i \cos\frac{\theta_{i+1}}{2} + \Omega_{i-1}\frac{\eta_{i-1}  \sin\frac{\theta_i}{2} \sin\phi_{i-1} \sin\theta_{i-1}}{ \eta_i}\,,
\end{equation}
\begin{align}
    \dot{\phi}_i & =  2\Omega_i \cos\frac{\theta_{i+1}}{2} \cos\phi_i \cot\theta_i + 2 \Delta_i  - \Omega_{i-1}\frac{\cos\phi_{i-1} \sec\frac{\theta_{i}}{2} \sin\theta_{i-1} \eta_{i-1} }{2\eta_i}  \nonumber\\
    & - \Omega_i \frac{\cos\phi_i \sin\theta_i \sin \frac{\theta_{i+1}}{2} \eta_i \tan\frac{\theta_{i+1}}{2} }{2\eta_{i+1}} - \Omega_{i+1}\cos\frac{\theta_{i+2}}{2} \cos\phi_{i+1} \tan\frac{\theta_{i+1}}{2}\,.
\end{align}
The variational error rate also has a compact form
\begin{equation}
    \Gamma^2 = \frac{1}{K}\sum_{i=1}^{K} \Omega_i^2 \sin^2\frac{\theta_i}{2} \sin^2\frac{\theta_{i+1}}{2} \frac{\eta_i (1-\eta_i)}{\eta_{i+1}}\,.
\end{equation}

When $J\rightarrow \infty$, away from the south pole and the north pole, $\cos^{2J}\frac{\theta}{2}$ tends to zero very rapidly with increasing $J$. Taking this limit in Eq.~(\ref{eq:thetadot-maintext}) and (\ref{eq:phidot-maintext}), we obtain the simplified variational dynamics
\begin{equation}
    J\dot{\theta}_i \rightarrow \Omega_i \sin\phi_i \,,
\end{equation}
\begin{equation}
    J \dot{\phi}_i \rightarrow \Delta_i  + \Omega_i \cos\phi_i \cot\theta_i\,.
\end{equation}
These dynamics can also be derived from the semiclassical energy $E_{sc}=\sum_{i} -\Delta_i + \eta_i ( \Omega_i\sin\theta_i\cos\phi_i +\Delta_i (1-\cos\theta_i))$ through $\dot{\theta}_i = -\frac{1}{2\text{Im}G^c_{\phi_i,\theta_i}}\partial_{\phi_i} E_{sc}$ and $\dot{\phi}_i = -\frac{1}{2\text{Im}G^c_{\theta_i,\phi_i}}\partial_{\theta_i} E_{sc}$, where $-2\text{Im}G^c$ is the antisymmetric metric tensor. The variational dynamics are quite slow, as the spin operators are normalized by $J$. The variational error rate is given by
\begin{equation}
    \Gamma^2 \rightarrow \frac{1}{K}\sum_{i=1}^{K}\frac{\Omega_i^2 \eta_i (-1+2\eta_i)\cos^2\phi_i}{2J}\,.
\end{equation}
Interestingly, the scaling of $\Gamma$  depends on the parity of $K$, which can be justified from Eq.~(\ref{eq:eta-small-K}). This dependence can be rigorously analyzed by retaining the lowest-order terms in $x_i^2$ in $\Gamma$. For odd $K$, $\Gamma^2$ approaches to zero quickly as $\eta_i \rightarrow 1/2$ exponentially fast with  $J$. This implies that the timescale $\tau$ for the breakdown of quantum-classical correspondence, i.e. $\int_{0}^{\tau}\Gamma dt \gtrsim 1$, is exponential in $J$. For even $K$, we find that the time scale $\tau$  in general scales as $\sqrt{J}$. The difference in timescales between odd and even $K$ is an interesting feature, but we are unsure whether this feature is just a reflection on the limitation of our simple ansatz.

\section{Discussion}\label{sec:discussion}
In this work, we have used the time-dependent variational principle (TDVP) method to investigate the quantum dynamics of the spin-$J$ PXP model. This model is an effective model of the Rydberg atom array when the Rydberg blockade takes effect. To effectively capture the Rydberg physics and facilitate efficient analytical derivation, the minimally entangled MPS that satisfies the blockade requirement is adopted. Our ansatz differs from previous literature by permitting general $\mathbb{Z}_K$ discrete translational symmetry. By mathematical induction, we solve the transfer matrix over arbitrary interval of lattice sites. Combined with  the Perron-Frobenius theorem, it enables us to work out nearly all essential quantities in the variational dynamics and quantum leakage. The only remaining quantity to determine is the inverse of the connected Gram matrix. Fortunately, the inverse has a compact form, revealing that its matrix elements decay rapidly with distance from the diagonal. As a result, the variational dynamics and error rate can be efficiently truncated. 

Our study of the PXP model is based on the TDVP method. Recently, there is tremendous effort in investigating Rydberg physics with variational principle, e.g., TDVP in higher spatial dimension \cite{Pichler2024} and larger blockade radius  ansatz \cite{Serbyn2024-b}. The advantages of the variational method are twofold. First, it offers analytical tractability. The analytical nature of TDVP equations mitigates numerical instabilities during integration and provides a solid foundation for rigorous analysis of the variational dynamics. The analytical variational method is complementary to the tensor-network based numerical techniques \cite{Deng2023,Nora2024}. Second, TDVP dynamics presents a clear semi-classical picture, offering a novel paradigm for quantum-classical correspondence in the context of quantum many-body system \cite{Turner2021}. At the same time, there are obvious limitations in TDVP method. Most noticeable is the choice of an appropriate variational anstaz. While we find the $\mathbb{Z}_K$ transfer matrix through mathematical induction, we are not aware of a universal approach to solve $\mathbb{Z}_K$ transfer matrix for a broader class of variational ansatz, especially for large bond-dimensional case, which can be used to diagnose quantum chaos \cite{Hallam2019}.  

Besides the variational ansatz, there are several promising directions that worth further exploration in Rydberg atom arrays using variational method.
First, we can systematically construct the adiabatic path between the initial state and the target state using optimal control theory (by minimizing the quantum leakage with fixed boundary conditions), which may be useful for experimental preparation of GHZ state with high fidelity \cite{lukin2019-GHZ}. 
Second, it is interesting to investigate whether long-time oscillations for specific initial states are related with period orbits, which need not necessarily be unstable. Indeed, enhanced revival and suppressed entanglement growth have been identified for initial states on KAM tori \cite{Serbyn2020}. Nevertheless, the connection between quantum many-body scarring and periodic orbits in phase space remains incomplete.
Third, the quantum phase transition with competing crystalline order is still perplexing due to its non-perturbative feature. While the sublattice symmetry breaking has been studied experimentally \cite{Lukin2017} and numerically \cite{Subir2020}, a complete analytical understanding is still lacking. 
From the perspective of quantum-classical correspondence, the ground state corresponds to a fixed point in the phase space. Therefore the spatial periodicity of these fixed points may shed light on the sublattice symmetry of the ground state.  
Last but not the least, experiments have demonstrated that periodic detuning can enhance quantum many-body oscillations \cite{Lukin2021}, which  can be used for the construction of robust discrete time crystal. 
The stabilization mechanism of periodic driving might be understood by studying properties of Floquet fixed points in the variational dynamics.

In conclusion, while the TDVP method offers powerful tools for understanding non-equilibrium quantum dynamics, its limitations and the ongoing exploration of new variational ansatz provide exciting avenues for future research in quantum optimal control, quantum phase transitions, and quantum chaos.


\section*{Acknowledgments}
We thank Peter Zoller, W. V. Liu, and Hannes Pichler for stimulating discussions. 
This work was supported by the National Natural Science Foundation of China (Grants No.  92365202,  No. 12475011, and No.  11921005),  the  National Key R\&D Program of China (2024YFA1409002), 
and Shanghai Municipal Science and  Technology Major Project (Grant No.2019SHZDZX01). 
\appendix

\section{Spin coherent state}\label{sec:spin coherent state}
Let us briefly review the properties of the spin coherent state. We follow the convention in \cite{Gilmore1990}, in which the reference state is the ``spin down'' state $\ket{J,-J}$ at the south pole of $\mathbb{S}^2$. We will use $\ket{0}$ and $\ket{J,-J}$ interchangely. The spin coherent state is generated by rotation of the reference state $\ket{\Omega}\equiv \ket{\theta,\phi}=e^{\xi S^{+}-\xi^{*}S^{-}}|J,-J\rangle$, where $S^{\pm}=S^x\pm \ii S^y$ and $\xi=\frac{\theta}{2}e^{-\mathrm{i}\phi}$. The spin coherent states constitute  an over-complete basis with the resolution of identity $ 1=\frac{2J+1}{4\pi}\int \ket{\Omega}\bra{\Omega}d\Omega$, where $d\Omega=\sin\theta d\theta d\phi$. For every operator in the Hilbert space $\mathbb{C}^{2J+1}$, there is a diagonal form $A=\frac{2J+1}{4\pi}\int A^u\ket{\Omega}\bra{\Omega}d\Omega$, where $A^u$ is a smooth function of $\theta$ and $\phi$. Importantly, $A^u$ is dual to $A_l\equiv\bra{\Omega}A\ket{\Omega}$ in the sense of $\Tr(AB)=\frac{2J+1}{4\pi}\int A^u B_l\ket{\Omega}\bra{\Omega}d\Omega=\frac{2J+1}{4\pi}\int A_l B^u\ket{\Omega}\bra{\Omega}d\Omega$.

The Vaidman's formulae for spin operators and spin coherent states are 
\begin{align}
S^{+}\ket{\Omega} &=Je^{\mathrm{i}\phi}\sin\theta \ket{\Omega} +\sqrt{2J}\frac{1+\cos\theta}{2} \ket{\Omega}^{\perp}_{S^{+}} \label{P8eq:spin-vaidman1}\\
S^{-}\ket{\Omega} &= Je^{-\mathrm{i}\phi}\sin\theta \ket{\Omega} + \sqrt{2J}e^{-2\ii \phi}\frac{\cos\theta-1}{2} \ket{\Omega}^{\perp}_{S^{-}} \label{P8eq:spin-vaidman2}\\
S^{z}\ket{\Omega} &= -J\cos\theta \ket{\Omega} + \sqrt{2J}\frac{e^{-\ii \phi}\sin\theta}{2} \ket{\Omega}^{\perp}_{S^{z}},
\label{P8eq:spin-vaidman3}
\end{align}
where these orthogonal states happen to be the same  $\ket{\Omega}^{\perp}_{S^{+}}=\ket{\Omega}^{\perp}_{S^{-}}=\ket{\Omega}^{\perp}_{S^{z}}=e^{\xi S^{+}-\xi^{*}S^{-}}\ket{J,-J+1}$. The absolute values of coefficients of orthogonal states are the standard deviation of spin operators. Remind of an important property of coherent state
\begin{equation}
    S^{+}\ket{\theta,\phi} = \frac{1}{\tau} (J+S^{z})\ket{\theta,\phi},\quad S^{-}\ket{\theta,\phi} = \tau (J-S^{z})\ket{\theta,\phi},
    \label{eq:coherent-property}
\end{equation}
where $\tau = \tan\frac{\theta}{2}e^{-\ii \phi}$.
It is a direct corollary from Vaidman's formulae (\ref{P8eq:spin-vaidman1})-(\ref{P8eq:spin-vaidman3}). 
The derivative of the spin coherent state is 
\begin{align}
    \partial_{\mu} \ket{\theta,\phi} 
     = -J \frac{\partial_{\mu} \abs{\tau}^2}{1+\abs{\tau}^2}\ket{\theta,\phi} + (\partial_{\mu}\tau) S^{+}\ket{\theta,\phi}
     \equiv B_{\mu} \ket{\theta,\phi},
     \label{eq:def-Bmu}
\end{align}
where we use 
$\ket{\theta,\phi} = (1+|\tau|^{2})^{-J}e^{\tau S^{+}}|J,-J\rangle $. According to Eq.~(\ref{eq:coherent-property}), the expression of $B_{\mu}$ is not unique.  We can choose 
a specific form
\begin{align}
    B_{\mu} &= J \frac{\partial_{\mu}\tau}{\tau}-J\frac{\partial_{\mu} \abs{\tau}^2}{1+\abs{\tau}^2}+\frac{\partial_{\mu}\tau}{\tau}S^{z} \nonumber\\
    & = J(\cot\theta \partial_{\mu}\theta -\ii \partial_{\mu}\phi)+ J(\frac{\partial_{\mu}\theta}{\sin\theta}-\ii \partial_{\mu}\phi) s^{z},
    \label{eq:Bmu-sz}
\end{align}
where $s^z=J^{-1}S^z$ is normalized spin-$J$ operators. The advantage is that now $B_{\mu}$ commutes with the projector operator $P=\ket{0}\bra{0}$ of unexcited state. 

We will frequently encounter expectation values of operators in the coherent state. In that case, the generating functions derived by Gilmore become very useful
\begin{align}
    &\bra{\theta,\phi} e^{\alpha_{\_}S^{-}} e^{\alpha_{0}S^{z}} e^{\alpha_{+}S^{+}}\ket{\theta,\phi} \nonumber\\
    &= (1+\abs{\tau}^2)^{-2J} \Big(e^{-\frac{1}{2}\alpha_0}+e^{\frac{1}{2}\alpha_0} (\tau^{*}+\alpha_{\_})(\tau + \alpha_{+}) \Big)^{2J}
\end{align}
\begin{align}
    &\bra{\theta,\phi} e^{\alpha_{+}S^{+}} e^{\alpha_{0}S^{z}} e^{\alpha_{\_}S^{-}}\ket{\theta,\phi} \nonumber\\
    & = (1+\abs{\tau}^2)^{-2J} \Big(e^{\frac{1}{2}\alpha_0} \abs{\tau}^2 +e^{-\frac{1}{2}\alpha_0} (\alpha_{+}\tau^{*}+1)(\tau \alpha_{\_} + 1) \Big)^{2J}.
\end{align}
By taking derivative of $\alpha$ in the generating function and finally setting $\alpha=0$, we can solve any $k$-point correlation function of $S^{\pm}$ and $S^z$. For convenience, we will list the expectation $\bra{\Omega}\mathcal{O}\ket{\Omega}$, $\bra{0}\mathcal{O}\ket{\Omega}$ and $\bra{0}\mathcal{O}\ket{0}$ in Table~\ref{tab:one-site op} and Table~\ref{tab:2-site op}. The spin operator is normalized $s^{\alpha}=J^{-1} S^{\alpha}$. To simplify the notation, we present results in terms of $\tau = \tan\frac{\theta}{2}e^{-\ii \phi}$ and $x \equiv \inner{0}{\theta,\phi}= \cos^{2J}\frac{\theta}{2}$.
\begin{table}[h]
    \centering
    \begin{tabular}{|c|c|c|c|c|c|}
    \hline
         $\mathcal{O}$ &   $P$&$s^z$&  $s^+$&   $s^-$ &$B_{\mu}$ \\
         \hline\hline
         $\bra{\Omega}\mathcal{O}\ket{\Omega}$ &    $x^2$&$-\cos\theta$&   $\sin\theta e^{\mathrm{i}\phi}$&  $\sin\theta e^{-\mathrm{i}\phi}$ &$\mathrm{i} J (\cos \theta -1) \partial_{\mu}\phi  $ \\
         \hline
         $\bra{0}\mathcal{O}\ket{\Omega}$ &    $x$&$-x$&   $0$&   $2x \tau $ &$-Jx\tan\frac{\theta}{2} \partial_{\mu}\theta$ \\
         \hline
 $\bra{0}\mathcal{O}\ket{0}$&  $1$&$-1$& $0$& $0$&$-J\tan\frac{\theta}{2} \partial_{\mu}\theta$ \\
 \hline
 \end{tabular}
    \caption{Expectation value of single onsite operator}
    \label{tab:one-site op}
\end{table}

\begin{table}[h]
    \centering
    \begin{tabular}{|c|>{\centering\arraybackslash}p{0.4\linewidth}|c|c|}
    \hline
       $\mathcal{O}$  & $\bra{\Omega}\mathcal{O}\ket{\Omega}$ & $\bra{0}\mathcal{O}\ket{\Omega}$  &$\bra{0}\mathcal{O}\ket{0}$\\
       \hline\hline
        $s^{x}s^{x}$ &  $\frac{1+(-1+2J)\cos^2\phi\sin^2\theta }{2J}$& $\frac{2J-1}{2J}x \tau^2 + \frac{x}{2J}$&$\frac{1}{2J}$\\
        \hline
       $s^{x}s^{z}$  &  $\frac{\sin\theta}{2J}  ((1-2 J) \cos\theta \cos\phi-\mathrm{i}\sin\phi)$& $\frac{-J+1}{J} x \tau$ &$0$\\
       \hline
         $s^z s^x$&  $\frac{\sin\theta}{2J}  ((1-2 J) \cos\theta \cos\phi +\mathrm{i}\sin\phi)$& $-x\tau$ &$0$\\
        \hline
         $s^z s^z$&  $\frac{1}{2J}  ((2 J-1) \cos^2\theta +1)$& $x$&$1$\\
         \hline
         $s^{x}B_{\mu}$ & 
             $\frac{\cos\theta\cos\phi-\ii \sin\phi}{2} \partial_{\mu}\theta + \frac{-\sin\phi+\ii \cos\phi (-2J+(-1+2J)\cos\theta)}{2} \sin\theta \partial_{\mu}\phi$
          & $-\ii x \tau \partial_{\mu}\phi + \frac{1-J+J\cos\theta}{\sin\theta} x\tau  \partial_{\mu}\theta $ & 0 \\
          \hline
         $s^{z} B_{\mu}$&  $\ii (-1+(-1+2J)\cos\theta)\sin^2\frac{\theta}{2}\partial_{\mu}\phi +\frac{\sin\theta}{2}\partial_{\mu}\theta$&  $J x \tan\frac{\theta}{2} \partial_{\mu}\theta$&$J \tan\frac{\theta}{2} \partial_{\mu}\theta$\\
         \hline
 $B^{\dagger}_{\mu}B_{\nu}$& $\frac{J}{2}\partial_{\mu}\theta \partial_{\nu}\theta -\ii \frac{J\sin\theta}{2}\partial_{\mu}\theta \partial_{\nu}\phi + \ii \frac{J\sin\theta}{2}\partial_{\mu}\phi \partial_{\nu}\theta + J \frac{1+6J-8J\cos\theta +(-1+2J)\cos2\theta}{4} \partial_{\mu}\phi \partial_{\nu}\phi$& $J^2 x\tan^2\frac{\theta}{2}\partial_{\mu}\theta \partial_{\nu}\theta$ & $J^2 \tan^2\frac{\theta}{2}\partial_{\mu}\theta \partial_{\nu}\theta$\\
 \hline
    \end{tabular}
    \caption{Correlation function of two onsite operators}
    \label{tab:2-site op}
\end{table}

\section{Properties of the transfer matrices}\label{sec:transfer matrix}
The transfer matrix for a unit cell is defined as the product of single-site transfer matrices
\begin{equation}
    T_{[1,K]} \equiv T_1\cdot T_2\cdots T_K,
\end{equation}
where $T_i$ is the transfer matrix at site $i$
\begin{equation}
    T_i = \sum_{\sigma} (A^{\sigma})^{*}\otimes A^{\sigma} = \left(\begin{array}{cccc}
x_i^{2} & 0 & 0 & 1-x_i^{2}\\
x_i & 0 & 0 & 0\\
x_i & 0 & 0 & 0\\
1 & 0 & 0 & 0
\end{array}\right).
\end{equation}
We will solve $T_{[1,K]}$ through mathematical induction.  Suppose that $T_{[1,K-1]}\equiv \prod_{i=1}^{K-1}T_i$ has the form
\[
T_{[1,K-1]} =  \left(\begin{array}{cccc}
a_{K-1} & 0 & 0 & b_{K-1}\\
c_{K-1} & 0 & 0 & d_{K-1}\\
c_{K-1} & 0 & 0 & d_{K-1}\\
e_{K-1} & 0 & 0 & f_{K-1}
\end{array}\right).
\]
Apparently $T_{[1,1]}=T_1$ is of the form with 
\[
a_1=x_1^2,\quad b_1 = 1- x_1^2,\quad c_1=x_1,\quad e_1=1,\quad d_1 = f_1 = 0.
\]
It is easy to check that $T_{[1,K]} = T_{[1,K-1]}\cdot T_K$ has the same form as $T_{[1,K-1]}$.
The recursion formulae for the pair $(a,b)$ are
\begin{align*}
a_K &= a_{K-1} + (-1+x_{K}^2)a_{K-1} -(-1+x_{K-1}^2)a_{K-2}\\
b_{K} &= a_{K-1}(1-x_K^2).
\end{align*}
The relations also hold for $(c,d)$ and $(e,f)$ separately. It is useful to define 
\[
a_0 =1,\quad b_0 = 0,\quad c_0 = 0,\quad d_0 = x_1,\quad e_0 = 0,\quad f_0 = 1 
\]
such that the recursion relations hold even when $K=1$. 
We find, by 
iteratively replacing the first term $a_i$ in the right hand side of the recursion relation $a_{i+1}=a_i + (-1+x_{i+1}^2)a_i - (-1+x_{i}^2)a_{i-1}$, the last term $-(-1+x_{i}^2)a_{i-1}$ cancels with the term appearing in the substitution of $a_i$. The procedure ends at $i=2$.
\begin{align*}
a_K &= a_1 - (-1+x_1^2)a_0 + (-1+x_K^2)a_{K-1}\\
&= (a_1 - (-1+x_1^2)a_0)(1+ \sum_{m=2}^{K}\prod_{i=m}^{K}(-1+x_i^2)) \\
&\quad + a_0 \prod_{i=1}^{K}(-1+x_i^2).
\end{align*}
$b_K$ can be obtained from $a_{K-1}$
\[
b_K = -(a_1 - (-1+x_1^2)a_0)\sum_{m=2}^{K}\prod_{i=m}^{K}(-1+x_i^2) - a_0 \prod_{i=1}^{K}(-1+x_i^2).
\]
The solutions for other pairs $(c,d)$ and $(e,f)$ are the same if you replace $c (e)$ with $a$ and $d (f)$ with $b$. In terms of 
\begin{equation}
    \alpha_K = \sum_{m=2}^{K}\prod_{i=m}^{K}(-1+x_i^2), \quad \beta_K = \prod_{i=1}^{K}(-1+x_i^2)
\end{equation}
$T_{[1,K]}$ can be written as
\begin{equation}
T_{[1,K]} =  \left(\begin{array}{cccc}
1+ \alpha_K + \beta_K & 0 & 0 & -\alpha_K - \beta_K \\
x_1 (1+\alpha_K) & 0 & 0 & -x_1 \alpha_K \\
x_1 (1+\alpha_K) & 0 & 0 & -x_1 \alpha_K \\
1+\alpha_K & 0 & 0 & -\alpha_K
\end{array}\right).
\end{equation}
We see that $T_{[1,K]}$ is consistent with $T_1$ when $K=1$ since $\alpha_1=0$. 
The nonzero eigenvalues of  $T_{[1,K]}$ are 
\begin{equation}
    \lambda_1 = 1,\quad \lambda_2 = \beta_K 
\end{equation}
with the corresponding eigenvectors
\begin{align*}
    |R_1)&=(1,x_1,x_1,1)^{\text{T}}, \\
    (L_1|&=(\frac{1+\alpha_K}{1-\beta_K},0,0,1-\frac{1+\alpha_K}{1-\beta_K} ); \\
     |R_2)&= (1-\frac{1-\beta_K}{1+\alpha_K},x_1,x_1,1)^{\text{T}}, \\ 
     (L_2| &= (-\frac{1+\alpha_K}{1-\beta_K},0,0,\frac{1+\alpha_K}{1-\beta_K}).
\end{align*}
The null eigenvectors ($\lambda=0$) are
\begin{align*}
    |R_3)&=(0,0,1,0)^{\text{T}}, \quad (L_3|=(0,0,1,-x_1 ); \\
     |R_4)&= (0,1,0,0)^{\text{T}}, \quad (L_4| = (0,1,0,-x_1).
\end{align*}
The dominating eigenvalue is $\lambda_1=1$, since $\abs{\lambda_2}<1$ unless all $\theta_i=\pi$. The Perron-Frobenius theorem tells us
\begin{equation*}
    \lim_{m\rightarrow\infty}T_{[1,K]}^{m} = |R_1)(L_1|.
\end{equation*} 
With these eigenvectors, one finds
\begin{align}
    &(L_1|T_1\cdot T_2\cdots T_{i-1} \nonumber\\
    &=  (1+\frac{1+\alpha_K}{1-\beta_K}\beta_{i-1}+\alpha_{i-1},0,0,-\frac{1+\alpha_K}{1-\beta_K}\beta_{i-1}-\alpha_{i-1} ),
    \label{P8eq:left-act}
\end{align}
\begin{equation}
    T_{i+1}\cdot T_{i+2}\cdots T_{K} |R_1) = (1,x_{i+1},x_{i+1},1)^{\text{T}}.
    \label{P8eq:right-act}
\end{equation}
Denote $\alpha^{\prime}_i=\sum_{m=i+1}^{K}\prod_{j=m}^{K}(-1+x_j^2)$.  Eq.~(\ref{P8eq:left-act}) becomes
\begin{align}
    &(L_1|T_1\cdot T_2\cdots T_{i-1} \nonumber\\
    & = (\frac{1+\alpha_{i-1}+(1+\alpha^{\prime}_i)\beta_{i-1}}{1-\beta_K},0,0,1-\frac{1+\alpha_{i-1}+(1+\alpha^{\prime}_i)\beta_{i-1}}{1-\beta_K}),
    \label{P8eq:left-act2}
\end{align}
where for consistence we set 
\begin{equation}
    \alpha_0=-1,\quad \alpha_1 = 0,\quad \alpha^{\prime}_{K}=0,\quad \alpha^{\prime}_{K+1}=-1,\quad \beta_0=1.
\end{equation}
 For symbolic simplicity, we adopt the notation
\begin{equation}
    \eta_i \equiv \frac{1+\alpha_{i-1}+(1+\alpha^{\prime}_i)\beta_{i-1}}{1-\beta_K},
    \label{P8eq:etadef1}
\end{equation}
or $\eta_i = 1+\eta_1 \beta_{i-1} +\alpha_{i-1}$. If we further assume that the transfer matrix has period $K$, Eq.~(\ref{P8eq:etadef1}) can be further simplified as
\begin{equation}
    \eta_i = \frac{1+\sum_{m=i-K+1}^{i-1}\prod_{j=m}^{i-1}(-1+x_j^2)}{1-\beta_K},
    \label{P8eq:etadef2}
\end{equation}
where the index should be understood as modulo $K$. We can see that $\eta_i$ inherits the periodicity as  $\eta_{1}=\eta_{K+1}= \frac{1+\alpha_K}{1-\beta_K}$.

These results can be generalized directly to products of transfer matrix over an interval  $T_{[i,j]}\equiv T_{i}\cdot T_{i+1}\cdots T_{j}$
\begin{equation}
    T_{[i,j]}=\left(\begin{array}{cccc}
1+\alpha_{[i,j]}+\beta_{[i,j]} & 0 & 0 & -\alpha_{[i,j]}-\beta_{[i,j]}\\
x_{i} (1+\alpha_{[i,j]}) & 0 & 0 & -x_{i}\alpha_{[i,j]} \\
x_{i} (1+\alpha_{[i,j]}) & 0 & 0 & -x_{i}\alpha_{[i,j]} \\
1+\alpha_{[i,j]} & 0 & 0 & -\alpha_{[i,j]}
\end{array}\right),
\label{eq:Tij-def}
\end{equation}
with the parameters
\begin{equation}
    \alpha_{[i,j]} = \sum_{m=i}^{j}\prod_{k=m}^{j} (-1+x_k^2)-\prod_{k=i}^{j}(-1+x_k^2),
    \label{eq:alphaij}
\end{equation}
and
\begin{equation}
    \beta_{[i,j]} = \prod_{k=i}^{j}(-1+x_k^2).
    \label{eq:betaij}
\end{equation}
The eigenvalues are $\{1,\beta_{[i,j]},0,0\}$. The corresponding eigenvectors are
\begin{align*}
    |R_1)&=(1,x_i,x_i,1)^{\text{T}}, \\
    (L_1|&=(\frac{1+ \alpha_{[i,j]}}{1- \beta_{[i,j]}},0,0,1-\frac{1+ \alpha_{[i,j]}}{1- \beta_{[i,j]}}); \\
     |R_2)&= (\frac{ \alpha_{[i,j]}+ \beta_{[i,j]} }{1+ \alpha_{[i,j]}},x_i,x_i,1)^{\text{T}},\\
     (L_2| &= (-\frac{1+ \alpha_{[i,j]}}{1- \beta_{[i,j]}},0,0, \frac{1+ \alpha_{[i,j]}}{1- \beta_{[i,j]}}); \\
     |R_3)&=(0,0,1,0)^{\text{T}}, \quad (L_3|=(0,0,1,-x_i ); \\
     |R_4)&= (0,1,0,0)^{\text{T}}, \quad (L_4| = (0,1,0,-x_i). 
\end{align*}
One can immediately see from Eq.~(\ref{P8eq:etadef2}) if $K$ is the period of the transfer matrix
\begin{equation}
    \eta_i = \frac{1+\alpha_{[i,i+K-1]}}{1-\beta_{[i,i+K-1]}}=1+\frac{\sum_{m=i}^{i+K-1}\prod_{j=m}^{i+K-1}(-1+x_j^2)}{1-\prod_{j=i}^{i+K-1}(-1+x_j^2)}.
    \label{eq:etai}
\end{equation}
It is direct to verify that $\eta_i$ is periodic in $K$, i.e. $\eta_{i+K} = \eta_{i}$. 
From the definition Eq.~(\ref{eq:etai}) of $\eta$, the dominant right and left eigenvectors of $T_{[i,i+K-1]}$ can be written as 
\begin{equation}
    |r_i)\equiv (1,x_i,x_i,1)^{\text{T}}, \quad (l_i| \equiv (\eta_i,0,0,1-\eta_i).
    \label{eq:rili-def}
\end{equation}

From Eq.~(\ref{eq:alphaij}), (\ref{eq:betaij}) and (\ref{eq:etai}), we find
\begin{equation}
    \eta_{j+1}=\eta_i \beta_{[i,j]} +1+\alpha_{[i,j]}.
    \label{eq:eta-relation}
\end{equation}
This formula establishes the relation between different $\eta_i$. When $j=i$ the formula is simplified as
\begin{equation}
    \eta_{i+1} = 1+ (-1+x_i^2)\eta_i.
\end{equation}
Using Eq.~(\ref{eq:eta-relation}), we obtain the reduction formula
\begin{equation}
    (\eta_i,0,0,1-\eta_i)T_{[i,j]}=(\eta_{j+1},0,0,1-\eta_{j+1}),
    \label{eq:reduction1}
\end{equation}
Another reduction formula can be seen from Eq.~(\ref{eq:Tij-def})
\begin{equation}
    T_{[i,j]} (1,x_{j+1},x_{j+1},1)^{\text{T}} = (1,x_i,x_i,1)^{\text{T}}.
    \label{eq:reduction2}
\end{equation}

Besides the product of transfer matrices, we also encounter the transfer matrix with onsite operator and the partial derivative. For notation simplicity, we will omit the site index when no confusion will be made. When we insert an onsite operator, the transfer matrix becomes
\begin{align}
    T_{q} &\equiv  \sum_{\sigma,\sigma'} (A^{\sigma})^{*}\otimes q^{\sigma,\sigma'} A^{\sigma'}  \nonumber\\
    &=\scriptsize \left(\begin{array}{cccc}
\langle\theta,\phi|P\hat{q}P|\theta,\phi\rangle & \langle\theta,\phi|P\hat{q}Q|\theta,\phi\rangle & \langle\theta,\phi|Q\hat{q}P|\theta,\phi\rangle & \langle\theta,\phi|Q\hat{q}Q|\theta,\phi\rangle\\
\langle\theta,\phi|P\hat{q}|0\rangle & 0 & \langle\theta,\phi|Q\hat{q}|0\rangle & 0\\
\langle0|\hat{q}P|\theta,\phi\rangle & \langle0|\hat{q}Q|\theta,\phi\rangle & 0 & 0\\
\langle0|\hat{q}|0\rangle & 0 & 0 & 0
\end{array}\right)
\label{eq:tran-mat-op}
\end{align}
where $Q=I-P$. Note that we use the hat notation to emphasize $\hat{q}$ is an operator. 
In the presence of the partial derivative, the transfer matrices are modified as
\begin{align}
    \partial_{\mu}T &\equiv \sum_{\sigma} (A^{\sigma})^{*}\otimes \partial_{\mu}A^{\sigma} \nonumber\\
    &  = \scriptsize\left(\begin{array}{cccc}
\langle\theta,\phi|PB_{\mu}P|\theta,\phi\rangle & \langle\theta,\phi|PB_{\mu}Q|\theta,\phi\rangle & \langle\theta,\phi|QB_{\mu}P|\theta,\phi\rangle & \langle\theta,\phi|QB_{\mu}Q|\theta,\phi\rangle\\
0 & 0 & 0 & 0\\
\langle0|B_{\mu}P|\theta,\phi\rangle & \langle0|B_{\mu} Q|\theta,\phi\rangle & 0 & 0\\
0 & 0 & 0 & 0
\end{array}\right),
\label{eq:muT}
\end{align}

\begin{align}
    \bar{\partial}_{\mu}T &\equiv \sum_{\sigma} (\partial_{\mu}A^{\sigma})^{*}\otimes \partial_{\mu}A^{\sigma} \nonumber\\
    &= \scriptsize \left(\begin{array}{cccc}
\langle\theta,\phi|P B_{\mu}^{\dagger} P|\theta,\phi\rangle & \langle\theta,\phi|P B_{\mu}^{\dagger} Q|\theta,\phi\rangle & \langle\theta,\phi|Q B_{\mu}^{\dagger} P|\theta,\phi\rangle & \langle\theta,\phi|Q B_{\mu}^{\dagger} Q|\theta,\phi\rangle\\
\langle\theta,\phi|P B_{\mu}^{\dagger}|0\rangle & 0 & \langle\theta,\phi|Q B_{\mu}^{\dagger}|0\rangle & 0\\
0 & 0 & 0 & 0\\
0 & 0 & 0 & 0
\end{array}\right),
\label{eq:barmuT}
\end{align}
and
\begin{align}
    \bar{\partial}_{\mu}\partial_{\nu}T_i &\equiv \sum_{\sigma} (\partial_{\mu}A^{\sigma})^{*}\otimes \partial_{\nu}A^{\sigma}\nonumber\\
    & = \scriptsize \left(\begin{array}{cccc}
x^2 \langle 0|B_{\mu}^{\dagger}B_{\nu}|0\rangle & x\langle0|B_{\mu}^{\dagger}B_{\nu}Q|\theta,\phi\rangle & x\langle\theta,\phi|QB_{\mu}^{\dagger}B_{\nu}|0\rangle & \langle\theta,\phi|QB_{\mu}^{\dagger}B_{\nu}Q|\theta,\phi\rangle\\
0 & 0 & 0 & 0\\
0 & 0 & 0 & 0\\
0 & 0 & 0 & 0
\end{array}\right),
\label{eq:barmunuT}
\end{align}
where we use the fact that the derivative of the spin coherent state is equivalent to the action of an operator on the same spin coherent state, i.e. Eq.~(\ref{eq:def-Bmu}). Note $P$ (and thus $Q$) commutes with partial derivative since $P$ (and thus $Q$) is independent of variational parameters.  Also $P$ (and thus $Q$) commutes with operators $B_{\mu}^{\dagger}$ and $B_{\mu}$. Meanwhile, there is a mixed type, where both partial derivative and local operator exist
\begin{align}
    \partial_{\mu}T_{q} &\equiv  \sum_{\sigma,\sigma'} (A^{\sigma})^{*}\otimes q^{\sigma,\sigma'} \partial_{\mu}A^{\sigma'}  \nonumber\\
    &= \scriptsize \left(\begin{array}{cccc}
\langle\theta,\phi|P\hat{q}B_{\mu}P|\theta,\phi\rangle & \langle\theta,\phi|P\hat{q}B_{\mu}Q|\theta,\phi\rangle & \langle\theta,\phi|Q\hat{q}B_{\mu}P|\theta,\phi\rangle & \langle\theta,\phi|Q\hat{q}B_{\mu}Q|\theta,\phi\rangle\\
0 & 0 & 0 & 0\\
\langle0|\hat{q}B_{\mu}P|\theta,\phi\rangle & \langle0|\hat{q}B_{\mu}Q|\theta,\phi\rangle & 0 & 0\\
0 & 0 & 0 & 0
\end{array}\right)
\label{eq:tran-mat-deri-op}
\end{align}

\section{Calculation of TDVP equations of motion}
In this section we will calculate the TDVP equations, which involve the Gram matrix and its inverse and the partial derivative of the energy expectation. The properties of transfer matrices in Appendix~\ref{sec:transfer matrix} will be used to simply these expressions.

\subsection{Gram matrix}\label{sec:gram}

We first calculate the connected Gram matrix $G_{\mu_i,\nu_{i+k}}^{c}\equiv \inner{\partial_{\mu_i}\Psi}{\partial_{\nu_{i+k}}\Psi}_c$. 
Using the notation from Eq.~(\ref{eq:muT}), (\ref{eq:barmuT}), and (\ref{eq:Tij-def}), and applying Perron-Frobenius theorem,  we obtain the following expression for $1\leq k\leq K-1$
\begin{align}
    G_{\mu_i,\nu_{i+k}}^{c}&=\frac{L}{K}\sum_{m\geq 0}(l_i| \bar{\partial}_{\mu}T_i \cdot T_{[i+1,i+k+mK-1]}\cdot \partial_{\nu}T_{i+k+mK} |r_{i+k+mK+1})_c \nonumber\\
    & + \frac{L}{K}\sum_{m\leq -1} (l_{i+k+mK}| \partial_{\nu}T_{i+k+mK} \cdot T_{[i+k+mK+1,i-1]} \cdot \bar{\partial}_{\mu}T_i |r_{i+1})_c.
    \label{eq:Gmunu}
\end{align}
The index $m$ runs over different unit cell. $(l_i|$ and $|r_i)$ are dominant left and right eigenvectors of $T_{[i,i+K-1]}$, respectively.  We define $T_{[i,j]}=I$ when $i=j+1$, which occurs for $k=1$ and $k=K-1$. The subscript ``c'' represents removal of disconnected parts $ (l_i| \bar{\partial}_{\mu}T_i |r_{i+1}) \cdot (l_{i+k+mK}|\partial_{\nu}T_{i+k+mK} |r_{i+k+mK+1})$ from $(l_i| \bar{\partial}_{\mu}T_i \cdot T_{[i+1,i+k+mK-1]}\cdot \partial_{\nu}T_{i+k+mK} |r_{i+k+mK+1})$ and $ (l_{i+k+mK}| \partial_{\nu}T_{i+k+mK} |r_{i+k+mK+1})\cdot (l_i|\bar{\partial}_{\mu}T_i |r_{i+1})$ from $(l_{i+k+mK}| \partial_{\nu}T_{i+k+mK} \cdot T_{[i+k+mK+1,i-1]} \cdot \bar{\partial}_{\mu}T_i |r_{i+1})$. From Eq.~(\ref{eq:muT}), (\ref{eq:barmuT}) and (\ref{eq:Tij-def}), we have
\begin{equation}
    (l_i| \partial_{\mu}T_i |r_{i+1}) = (l_i| \bar{\partial}_{\mu}T_i |r_{i+1})^{*} = \eta_i \bra{\theta_i,\phi_i} B_{\mu_{i}} \ket{\theta_i,\phi_i},
\end{equation}
and
\begin{align}
    & (l_i| \bar{\partial}_{\mu}T_i \cdot T_{[i+1,j-1]}\cdot \partial_{\nu}T_{j} |r_{j+1})_c \nonumber\\
    &\equiv (l_i| \bar{\partial}_{\mu}T_i \cdot T_{[i+1,j-1]}\cdot \partial_{\nu}T_{j} |r_{j+1}) - (l_i| \bar{\partial}_{\mu}T_i |r_{i+1}) \cdot (l_j| \partial_{\nu}T_{j} |r_{j+1})\nonumber\\
    & = \bra{\theta_j,\phi_j} B_{\nu_{j}} \ket{\theta_j,\phi_j} \eta_i \beta_{[i+1,j-1]} \Big( -\eta_{i+1} \bra{\theta_i,\phi_i} B_{\mu_{i}}^{\dagger} \ket{\theta_i,\phi_i} + x_i^2 \bra{0}_i B_{\mu_{i}}^{\dagger} \ket{0}_i \Big). 
    \label{eq:TmuTnu-connected}
\end{align}
Note that this result holds also for $j=i+1$, as long as we use the convention  $\beta_{[k,l]}=I$ when $k=l+1$.
Substitute Eq.~(\ref{eq:TmuTnu-connected}) into Eq.~(\ref{eq:Gmunu}), we have
\begin{align}
    G_{\mu_i,\nu_{i+k}}^{c}&= \frac{L}{K}\Big( \eta_i (-\eta_{i+1} \bra{\Omega_{i}} B_{\mu_{i}}^{\dagger} \ket{\Omega_{i}} + x_{i}^2 \bra{0}_i B_{\mu_{i}}^{\dagger} \ket{0}_i )\frac{\beta_{[i+1,i+k-1]}}{1-\beta_{[1,K]}} \bra{\Omega_{i+k}} B_{\nu_{i+k}} \ket{\Omega_{i+k}} \nonumber\\
    &+ \eta_{i+k}\frac{\beta_{[i+k-K+1,i-1]}}{1-\beta_{[1,K]}} \bra{\Omega_{i}} B_{\mu_{i}}^{\dagger} \ket{\Omega_{i}} (-\eta_{i+k+1} \bra{\Omega_{i+k}} B_{\nu_{i+k}} \ket{\Omega_{i+k}}\nonumber\\
    &\quad+x_{i+k}^2 \bra{0}_{i+k} B_{\nu_{i+k}} \ket{0}_{i+k} )  \Big),
\end{align}
where for notation simplicity we use $\ket{\Omega_i}\equiv \ket{\theta_i,\phi_i}$.   The $\frac{1}{1-\beta_{[1,K]}}$ comes from the summation $\sum_n\beta_{[1,K]}^n=1+\beta_{[1,K]} +\beta_{[1,K]}^2+ \cdots$.

We now deal with $k=0$ case. Similarly
\begin{align}
    G_{\mu_i,\nu_{i}}^{c} & =\frac{L}{K}  (l_i| \bar{\partial}_{\mu}\partial_{\nu}T_i|r_{i+1})\nonumber\\
    &+\frac{L}{K}\sum_{m\geq 1}(l_i| \bar{\partial}_{\mu}T_i \cdot T_{[i+1,i+mK-1]}\cdot \partial_{\nu}T_{i+mK} |r_{i+mK+1}) \nonumber\\
    & + \frac{L}{K}\sum_{m\leq -1} (l_{i+mK}| \partial_{\nu}T_{i+mK} \cdot T_{[i+mK+1,i-1]} \cdot \bar{\partial}_{\mu}T_i |r_{i+1}) \nonumber\\
    & - \frac{L}{K}\sum_{m}  (l_i| \bar{\partial}_{\mu}T_i|r_{i+1})(l_i|\partial_{\nu}T_i|r_{i+1}). 
\end{align}
From Eq.~(\ref{eq:TmuTnu-connected}) and (\ref{eq:barmunuT}), we obtain
\begin{align}
    G_{\mu_i,\nu_{i}}^{c}&= \frac{L}{K}\Big( \eta_i \Big( \bra{\Omega_{i}} B_{\mu_{i}}^{\dagger}B_{\nu_{i}} \ket{\Omega_{i}} - \eta_i \bra{\Omega_{i}} B_{\mu_{i}}^{\dagger} \ket{\Omega_{i}} \bra{\Omega_{i}} B_{\nu_{i}} \ket{\Omega_{i}} \Big) \nonumber\\
    & + \eta_i \Big(-\eta_{i+1} \bra{\Omega_{i}} B_{\mu_{i}}^{\dagger} \ket{\Omega_{i}} + x_{i}^2 \bra{0}_i B_{\mu_{i}}^{\dagger} \ket{0}_i  \Big) \frac{\beta_{[i+1,i+K-1]}}{1-\beta_{[1,K]}} \bra{\Omega_{i}} B_{\nu_{i}} \ket{\Omega_{i}}\nonumber\\
    &+\eta_{i} \Big(-\eta_{i+1} \bra{\Omega_{i}} B_{\nu_{i}} \ket{\Omega_{i}} +x_{i}^2 \bra{0}_i B_{\nu_{i}} \ket{0}_i  \Big) \frac{\beta_{[i-K+1,i-1]}}{1-\beta_{[1,K]}} \bra{\Omega_{i}} B_{\mu_{i}}^{\dagger} \ket{\Omega_{i}} \Big).
    \label{eq:Gelem}
\end{align}
Substituting Table \ref{tab:one-site op} and \ref{tab:2-site op} into the components of the $2K\times 2K$ connected Gram matrix Eq.~(\ref{eq:Gelem}), we have 
for $1\leq k \leq K-1$
\begin{equation}
    G_{\theta_i,\theta_{i+k}}^{c} = 0,
\end{equation}
\begin{equation}
    G_{\theta_i,\phi_{i+k}}^{c} = \frac{L}{K}\Big( \ii J^2 \eta_i x_i^2 \tan\frac{\theta_i}{2} \frac{\beta_{[i+1,i+k-1]}}{1-\beta_{[1,K]}} (1-\cos\theta_{i+k})\Big),
\end{equation}
\begin{equation}
    G_{\phi_i,\theta_{i+k}}^{c} =  \frac{L}{K}\Big(- \ii J^2 \eta_{i+k} x^2_{i+k} \tan\frac{\theta_{i+k}}{2} \frac{\beta_{[i+k-K+1,i-1]}}{1-\beta_{[1,K]}} (1-\cos\theta_{i}) \Big),
\end{equation}
\begin{align}
    G_{\phi_i,\phi_{i+k}}^{c} &=  \frac{L}{K}\Big( -J^2\eta_i \eta_{i+1} (1-\cos\theta_i) \frac{\beta_{[i+1,i+k-1]}}{1-\beta_{[1,K]}} (1-\cos\theta_{i+k}) \nonumber\\
    &- J^2\eta_{i+k} \eta_{i+k+1}  (1-\cos\theta_{i+k})\frac{\beta_{[i+k-K+1,i-1]}}{1-\beta_{[1,K]}} (1-\cos\theta_{i}) \Big).
\end{align}
For $k=0$
\begin{equation}
    G_{\theta_i,\theta_{i}}^{c} =  \frac{L}{K}\Big( \eta_i \frac{J}{2} \Big),
\end{equation}
\begin{equation}
    G_{\theta_i,\phi_{i}}^{c} =  \frac{L}{K}\Big( -\ii \eta_i J\frac{\sin\theta_i}{2} + \ii J^2 \eta_i x_i^2 \tan\frac{\theta_i}{2} \frac{\beta_{[i+1,i+K-1]}}{1-\beta_{[1,K]}} (1-\cos\theta_{i}) \Big),
\end{equation}
\begin{equation}
    G_{\phi_i,\theta_{i}}^{c} =  \frac{L}{K}\Big( \ii \eta_i J\frac{\sin\theta_i}{2} - \ii J^2 \eta_{i} x^2_{i} \tan\frac{\theta_{i}}{2} \frac{\beta_{[i-K+1,i-1]}}{1-\beta_{[1,K]}} (1-\cos\theta_{i}) \Big),
\end{equation}
\begin{align}
    G_{\phi_i,\phi_{i}}^{c} &=  \frac{L}{K}\Big( \eta_i(1-\eta_i)J^2 (1-\cos\theta_i)^2+ \eta_i J \frac{\sin^2\theta_i}{2} \nonumber\\
    & -2 J^2\eta_i \eta_{i+1} (1-\cos\theta_i) \frac{\beta_{[i+1,i+K-1]}}{1-\beta_{[1,K]}} (1-\cos\theta_{i}) \Big).
\end{align}

\subsection{Inverse of the Gram matrix}\label{sec:inv-gram}
To solve the equations of motion, we need to calculate the inverse of the Gram matrix $(\text{Im} G^{c}_{\bm{\theta,\bm{\phi}}})^{-1}$. Inverting a $K\times K$ matrix is generally a challenging task. However, we notice that  $\text{Im} G^{c}_{\bm{\theta,\bm{\phi}}}$ is closely related to the matrix $A$
\begin{align}
    A_{ij}=\begin{cases}
a_i b_i \,\prod_{m=i+1}^{i+K-1}z_m & \text{if}\,i=j,\\ 
a_i b_j \,\prod_{m=i+1}^{j-1}z_m & \text{if}\, i<j,\\ 
a_i b_j \,\prod_{m=i+1}^{j+K-1}z_m & \text{if}\, i>j,
\end{cases}
\label{eq:Amat}
\end{align}
where $A$ has an analytically tractable inverse given by 
\begin{align}
    A^{-1}_{ij}=\begin{cases}
-\frac{z_i}{a_i b_i(1-\prod_{m=1}^{K} z_m)} & \text{if}\, i=j,\\
\frac{1}{a_{i-1} b_j (1-\prod_{m=1}^{K} z_m)} & \text{if}\,j=i-1,\\
0 & \text{otherwise.}
\end{cases}
\label{eq:Ainvmat}
\end{align}
The matrix $A$ closely approximates $G^{c}_{\bm{\theta},\bm{\phi}}$, differing only in the diagonal term, where $-\ii \eta_i J\frac{\sin\theta_i}{2}$ is added. This difference introduces a diagonal correction matrix $C$ with elements $C_{ij}=c_{i}\delta_{ij}$, allowing us to rewrite the problem as inverting $A+C$, which is
\begin{equation}
    (A+C)^{-1}=C^{-1}-C^{-1}(A^{-1}+C^{-1})^{-1}C^{-1}.
\end{equation}
We note that, after factoring out the common term $1-\prod_{m=1}^{K} z_m$, its structure resembles that of $A^{-1}$.
The inverse can be obtained analogously from Eq.~(\ref{eq:Amat}) and (\ref{eq:Ainvmat}), yielding  
\begin{align}
    (A^{-1}+C^{-1})^{-1}_{ij} = \begin{cases}
    \frac{1-\prod_{m=1}^{K} z_m}{1-\prod_{m=1}^{K} \Tilde{c}_m} a_i b_i  \prod_{m=i+1}^{i+K-1}\Tilde{c}_m & \text{if}\,i=j,\\
    \frac{1-\prod_{m=1}^{K} z_m}{1-\prod_{m=1}^{K} \Tilde{c}_m} a_i b_j  \prod_{m=i+1}^{j-1}\Tilde{c}_m & \text{if}\, i<j,\\
    \frac{1-\prod_{m=1}^{K} z_m}{1-\prod_{m=1}^{K} \Tilde{c}_m} a_i b_j  \prod_{m=i+1}^{j+K-1}\Tilde{c}_m & \text{if}\, i>j,
    \end{cases}
\end{align}
where $\Tilde{c}_i=z_i - \frac{a_i b_i}{c_i}$. 
Finally, the inverse of $A+C$ is given by
\begin{align}
    (A+C)^{-1}_{ij} = \begin{cases}
    c_i^{-1}-c_i^{-2}\frac{1-\prod_{m=1}^{K} z_m}{1-\prod_{m=1}^{K} \Tilde{c}_m} a_i b_i  \prod_{m=i+1}^{i+K-1}\Tilde{c}_m & \text{if}\,i=j,\\
    -c_i^{-1} c_j^{-1} \frac{1-\prod_{m=1}^{K} z_m}{1-\prod_{m=1}^{K} \Tilde{c}_m} a_i b_j  \prod_{m=i+1}^{j-1}\Tilde{c}_m & \text{if}\,i<j,\\
    -c_i^{-1} c_j^{-1} \frac{1-\prod_{m=1}^{K} z_m}{1-\prod_{m=1}^{K} \Tilde{c}_m} a_i b_j  \prod_{m=i+1}^{j+K-1}\Tilde{c}_m & \text{if}\,i>j.
    \end{cases}
    \label{eq:invA+C}
\end{align}
This provides a formal solution $(\text{Im} G^{c}_{\bm{\theta},\bm{\phi}})^{-1}$ in principle.
Comparing with the connected Gram matrix, we find the parameters
\begin{align}
    a_i &= \eta_i x_i^2 \bra{0}_i B_{\mu_{i}} \ket{0}_i = -J \eta_i x_i^2 \tan\frac{\theta_i}{2},\nonumber\\
    b_i &= \frac{-\ii \bra{\Omega_{i}} B_{\mu_{i}} \ket{\Omega_{i}} }{1-\beta_{[1,K]}} = - \frac{J (1-\cos\theta_i)}{1-\beta_{[1,K]}}, \nonumber\\
    c_i & = -\eta_i J \frac{\sin\theta_i}{2},\nonumber\\
    \Tilde{c}_i &= z_i - \frac{a_i b_i}{c_i} = -1+ (\frac{2J}{1-\beta_{[1,K]}} \tan^2\frac{\theta_i}{2}+1)x_i^2.
    \label{eq:parameter-in-invG}
\end{align}
Substitute them into Eq.~(\ref{eq:invA+C})
\begin{align}
    (\text{Im} G^{c}_{\bm{\theta,\bm{\phi}}})^{-1}_{ij} = \frac{K}{L} \begin{cases}
    -\frac{2}{J\eta_i\sin\theta_i}-\frac{4 x_i^2 \tan^2\frac{\theta_i}{2}}{\sin\theta_i \eta_i} \frac{\prod_{m=i+1}^{i+K-1}\Tilde{c}_m}{1-\prod_{m=1}^{K} \Tilde{c}_m} & \text{if}\,i=j,\\
    -\frac{4 x_i^2 \tan\frac{\theta_i}{2} \tan\frac{\theta_j}{2} }{\sin\theta_i \eta_j} \frac{\prod_{m=i+1}^{j-1}\Tilde{c}_m}{1-\prod_{m=1}^{K} \Tilde{c}_m}   &  \text{if}\,i<j,\\
    -\frac{4 x_i^2 \tan\frac{\theta_i}{2} \tan\frac{\theta_j}{2} }{\sin\theta_i \eta_j} \frac{\prod_{m=i+1}^{j+K-1}\Tilde{c}_m}{1-\prod_{m=1}^{K} \Tilde{c}_m}  &  \text{if}\,i>j.
    \end{cases}
\end{align}
According to the relation $ (G^{c}_{\bm{\theta,\bm{\phi}}})_{ij} =(G^{c}_{\bm{\phi,\bm{\theta}}})^{*}_{ji} =-(G^{c}_{\bm{\phi,\bm{\theta}}})_{ji} $, we have 
\begin{equation}
    (\text{Im} G^{c}_{\bm{\theta,\bm{\phi}}})^{-1}_{ij}=-(\text{Im} G^{c}_{\bm{\phi,\bm{\theta}}})^{-1}_{ji}.
    \label{eq:ImGc-transpose}
\end{equation}

When $K$ is sufficiently large, we can set $\beta_{[1,K]}=0$ and 
\begin{equation}
    \Tilde{c}_i =  -1+ (2J \tan^2\frac{\theta_i}{2}+1)x_i^2.
    \label{eq:ctilde}
\end{equation}
The expression of $\Tilde{c}_i$ is surprisingly simple for $J=\frac{1}{2}$
\begin{equation}
    \Tilde{c}_i = 0. 
\end{equation}
In $J=\frac{1}{2}$ case, the nonzero elements $(\text{Im} G^{c}_{\bm{\theta,\bm{\phi}}})^{-1}_{ij}$ are 
\begin{equation}
    (\text{Im} G^{c}_{\bm{\theta,\bm{\phi}}})^{-1}_{ii} = \frac{K}{L}\Big( -\frac{2}{J\eta_i\sin\theta_i} \Big),
\end{equation}
\begin{equation}
    (\text{Im} G^{c}_{\bm{\theta,\bm{\phi}}})^{-1}_{i,i+1} = \frac{K}{L}\Big( -\frac{4 x_i^2 \tan\frac{\theta_i}{2} \tan\frac{\theta_{i+1}}{2} }{\sin\theta_i \eta_{i+1}} \Big). 
\end{equation}

\subsection{$\bra{\Psi}H\ket{\partial_{\mu}\Psi}_c$}\label{sec:muH}
We begin by noting that the matrix elements of $(\partial_{\mu_i}A_i) A_{i+1}$ and $A_i (\partial_{\mu_{i+1}}A_{i+1})$ do not contain any $(I_i-P_i)\ket{\theta_i,\phi_i}\otimes (I_{i+1}-P_{i+1})\ket{\theta_{i+1},\phi_{i+1}}$ components. Mathematically, that is
\begin{equation}
     P_{i,i+1}(\partial_{\mu_i}A_i)A_{i+1}=(\partial_{\mu_i}A_i)A_{i+1}, \quad P_{i,i+1}A_{i}(\partial_{\mu_{i+1}}A_{i+1})=A_{i}(\partial_{\mu_{i+1}}A_{i+1}).
\end{equation}
The equation implies that the operation of the partial derivative commutes with the projector $\mathcal{P}=\prod_i P_{i,i+1}$. Since $\mathcal{P}\ket{\Psi}=\ket{\Psi}$, the projector $\mathcal{P}$ can be omitted when computing $\bra{\Psi}H\ket{\partial_{\mu}\Psi}_c$. Consequently, the problem reduces to calculating $\bra{\Psi}\Omega_i s^{x}_i + \Delta_i s^{z}_i \ket{\partial_{\mu}\Psi}_c$.

For the expectation value of single onsite operator $q_i$, it can be expressed as
\begin{equation}
    (l_i| T_{q_{i}} |r_{i+1}) = \eta_i  \tilde{h}_{q_i} +  \bra{0}_i q_i \ket{0}_i,
    \label{eq:avq}
\end{equation}
where 
\begin{align}
    \tilde{h}_{q_i} &\equiv  + (\bra{0}_i q_i Q\ket{\Omega_i} + \bra{\Omega_i}Qq_i\ket{0}_i) x_i (-1+x_{i+1})  \nonumber\\
    &+\bra{\Omega_i} q_i \ket{\Omega_i} - \bra{0}_i q_i \ket{0}_i.
    \label{eq:h-def}
\end{align}

The connected part of $(l_i| (\partial_{\mu}T_i) \cdot T_{[i+1,j-1]}\cdot q_{j} |r_{j+1})$ is given by
\begin{align}
    & (l_i| (\partial_{\mu}T_i) \cdot T_{[i+1,j-1]}\cdot T_{q_{j}} |r_{j+1})_c \nonumber\\
    &\equiv (l_i| (\partial_{\mu}T_i) \cdot T_{[i+1,j-1]}\cdot T_{q_{j}} |r_{j+1}) - (l_i| (\partial_{\mu}T_i) |r_{i+1}) \cdot (l_j| T_{q_{j}} |r_{j+1})\nonumber\\
    & = \eta_i \big(-\eta_{i+1} \bra{\Omega}_i B_{\mu_{i}} \ket{\Omega}_i + x_i^2 \bra{0}_i B_{\mu_{i}} \ket{0}_i \big) \beta_{[i+1,j-1]}  \tilde{h}_{q_j}.
\end{align}
If we define $\beta_{[k+1,k]}=1$, this equation will hold for $j\geq i+1$. Similarly for $j\leq i-2$, we have
\begin{align}
    & (l_j|T_{q_{j}}\cdot T_{[j+1,i-1]} \cdot (\partial_{\mu}T_i) |r_{i+1})_c \nonumber\\
    &\equiv (l_j|T_{q_{j}}\cdot T_{[j+1,i-1]} \cdot (\partial_{\mu}T_i) |r_{i+1}) - (l_i| (\partial_{\mu}T_i) |r_{i+1}) \cdot (l_j| T_{q_{j}} |r_{j+1})\nonumber\\
    & =  - \tilde{h}_{q_j} \eta_j \eta_{j+1} \beta_{[j+1,i-1]} \bra{\Omega}_i B_{\mu_{i}} \ket{\Omega}_i.
\end{align}
For the case $j=i-1$, the expression becomes
\begin{align}
    & (l_{i-1}|T_{q_{i-1}}\cdot (\partial_{\mu}T_i) |r_{i+1})_c \nonumber\\
    &\equiv (l_{i-1}|T_{q_{i-1}} \cdot (\partial_{\mu}T_i) |r_{i+1}) - (l_i| (\partial_{\mu}T_i) |r_{i+1}) \cdot (l_{i-1}| T_{q_{i-1}} |r_{i})\nonumber\\
    & =  - \tilde{h}_{q_{i-1}} \eta_{i-1} \eta_{i} \bra{\Omega}_i B_{\mu_{i}} \ket{\Omega}_i + \eta_{i-1} x_{i-1} x_i \bra{0}_i B_{\mu_{i}} \ket{0}_i \bra{\Omega}_{i-1}Q_{i-1}q_{i-1}\ket{0}_{i-1}.
\end{align}

The self-term contribution $j=i$ is, according to Eq.~(\ref{eq:tran-mat-deri-op})
\begin{align}
    & (l_{i}| \partial_{\mu}T_{q_{i}} |r_{i+1})_c \nonumber\\
    &\equiv (l_{i}| \partial_{\mu}T_{q_{i}} |r_{i+1}) - (l_i| (\partial_{\mu}T_i) |r_{i+1}) \cdot (l_{i}| T_{q_{i}} |r_{i+1})\nonumber\\
    & = \eta_i (\tilde{h}_{q_i B_{\mu_i}} + \bra{0}_i q_i B_{\mu_i} \ket{0}_i )  - (\eta_i\tilde{h}_{q_i}+\bra{0}_i q_i\ket{0}_i ) \eta_{i} \bra{\Omega}_i B_{\mu_{i}} \ket{\Omega}_i, 
\end{align}
where the shorthand notation $\tilde{h}_{q_i B_{\mu_i}} \equiv \bra{\Omega_i} q_i B_{\mu_i} \ket{\Omega_i} - \bra{0}_i q_i B_{\mu_i} \ket{0}_i + (\bra{0}_i q_i B_{\mu_i} Q\ket{\Omega_i} + \bra{\Omega_i}Qq_i B_{\mu_i}\ket{0}_i) x_i (-1+x_{i+1})$ comes from Eq.~(\ref{eq:h-def}).

Combining these results and referring to Tables \ref{tab:one-site op} and \ref{tab:2-site op}, we obtain the real part of $\bra{\partial_{\theta_i}\Psi}H\ket{\Psi}_c$
\begin{align}
    &\text{Re} \bra{\partial_{\theta_i}\Psi}H\ket{\Psi}_c \nonumber\\
    & = \frac{L}{K}\bigg( -J \Omega_{i-1} \tan\frac{\theta_i}{2} x_{i-1}^2 \tan\frac{\theta_{i-1}}{2}\cos\phi_{i-1} x_i \eta_{i-1}\nonumber\\
    &  - \eta_i \Omega_i \cos\phi_i \Big( \frac{1}{2}+  x_i^2 (-1+x_{i+1}) (1 -2J + \frac{4J-1}{2\cos^2\frac{\theta_i}{2}})  \Big) \bigg) \nonumber\\
    & - \sum_{j=i}^{i+K-1} (\text{Im} G^{c}_{\bm{\theta},\bm{\phi}})_{ij} \frac{\Omega_i\cos\phi_i (\sin\theta_i +2x_i^2(-1+x_{i+1})\tan\frac{\theta_i}{2}) +\Delta_i (1-\cos\theta_i)}{J(1-\cos\theta_i)},
\end{align}
which contains the term involving $\text{Im} G^{c}_{\bm{\theta},\bm{\phi}}$. The imaginary part of $\bra{\partial_{\theta_i}\Psi}H\ket{\Psi}_c$ is
\begin{align}
    &\text{Im} \bra{\partial_{\theta_i}\Psi}H\ket{\Psi}_c \nonumber\\
    & =  \frac{L}{K}\Big( J \Omega_{i-1} \tan\frac{\theta_{i}}{2} \tan\frac{\theta_{i-1}}{2}\sin\phi_{i-1}  x^2_{i-1}x_i \eta_{i-1} \nonumber\\
    &+\eta_i \Omega_i \frac{\sin\phi_i}{2} + \Omega_i \eta_i x^2_i (-1+x_{i+1}) \frac{1}{\sin\theta_i}  \tan\frac{\theta_i}{2} \sin\phi_i \Big).
\end{align}
Similarly, the real and imaginary parts of  $\bra{\partial_{\phi_i}\Psi}H\ket{\Psi}_c$ are
\begin{align}
    &\text{Re} \bra{\partial_{\phi_i}\Psi}H\ket{\Psi}_c \nonumber\\
    & = \frac{L}{K}\Big( -\eta_i \Omega_i \frac{\sin\theta_i  \sin\phi_i}{2} - \eta_i \Omega_i \sin\phi_i x_i^2 (-1+x_{i+1})\tan\frac{\theta_i}{2} \Big),
\end{align}
and
\begin{align}
    &\text{Im} \bra{\partial_{\phi_i}\Psi}H\ket{\Psi}_c  
    \nonumber\\
    &=  \sum_{j=i}^{i+K-1} G^{c}_{\phi_i,\phi_j}\frac{\Omega_j\cos\phi_j (\sin\theta_j +2x_j^2(-1+x_{j+1})\tan\frac{\theta_j}{2}) +\Delta_j (1-\cos\theta_j)}{J(1-\cos\theta_j)} \nonumber\\
    & - \frac{L}{K}\Big( \frac{\eta_i}{2} \Omega_i \cos\phi_i \Big( \sin\theta_i + (\cos\theta_i +2J(1-\cos\theta_i) ) 2x_i^2 (-1+x_{i+1})\tan\frac{\theta_i}{2}\Big) \Big).
    \label{P8eq:imdphiH}
\end{align}

\subsection{TDVP equations}\label{sec:tdvp-eom}
To account for the contributions from the term $\text{Im} G^{c}_{\bm{\theta},\bm{\phi}}$ in  $\text{Re} \bra{\partial_{\mu_j}\Psi}H\ket{\Psi}_c$, we define a residual term $R_{\phi_j}$ for $\text{Re} \bra{\partial_{\phi_j}\Psi}H\ket{\Psi}_c$, which excludes contributions involving $(\text{Im} G^{c}_{\bm{\theta},\bm{\phi}})_{jl}$. This residual term is given by
\begin{align}
    R_{\phi_j} \equiv \frac{L}{K}\Big( -\eta_j \frac{\tan\phi_j}{2} \Omega_j\cos\phi_j (\sin\theta_j +2x_j^2(-1+x_{j+1})\tan\frac{\theta_j}{2}) \Big).
    \label{eq:resi-phiH}
\end{align}
Using this definition, along with the TDVP equations of motion and Eq.~(\ref{eq:ImGc-transpose}), we can write the equation for $\dot{\theta}_i$ as
\begin{align}
\dot{\theta}_i &= -\sum_j(\text{Im} G^{c}_{\bm{\phi},\bm{\theta}})_{ij}^{-1} \text{Re}\bra{\partial_{\phi_j}\Psi}H\ket{\Psi}_c \nonumber\\
     &= \sum_j(\text{Im} G^{c}_{\bm{\theta},\bm{\phi}})_{ji}^{-1} R_{\phi_j}.
     \label{eq:eom-theta}
\end{align}
Similarly, we define a residual term $R_{\theta_j}$  for $\text{Re} \bra{\partial_{\theta_j}\Psi}H\ket{\Psi}_c$, which excludes contributions involving $(\text{Im} G^{c}_{\bm{\theta},\bm{\phi}})_{jl}$. The residual term is given by
\begin{align}
    R_{\theta_j} &\equiv \frac{L}{K}\Big( -J \Omega_{j-1} \tan\frac{\theta_j}{2} x_{j-1}^2 \tan\frac{\theta_{j-1}}{2}\cos\phi_{j-1} x_j \eta_{j-1} \nonumber\\
    & - \eta_j \Omega_j \cos\phi_j \Big( \frac{1}{2}+  x_j^2 (-1+x_{j+1}) (1 -2J + \frac{4J-1}{2\cos^2\frac{\theta_j}{2}})  \Big) \Big).
    \label{eq:resi-thetaH}
\end{align}
The equation of motion for $\dot{\phi}_i$ can be expressed as
\begin{align}
    \dot{\phi}_i & = -\sum_j (\text{Im} G^{c}_{\bm{\theta},\bm{\phi}})_{ij}^{-1} \text{Re} \bra{\partial_{\theta_j}\Psi}H\ket{\Psi}_c\nonumber\\
    & = \frac{\Omega_i\cos\phi_i (\sin\theta_i +2x_i^2(-1+x_{i+1})\tan\frac{\theta_i}{2}) +\Delta_i (1-\cos\theta_i)}{J(1-\cos\theta_i)}  \nonumber\\
    & - \sum_j (\text{Im} G^{c}_{\bm{\theta},\bm{\phi}})_{ij}^{-1} \cdot R_{\theta_j}.
    \label{eq:eom-phi}
\end{align}

For the specific case of spin $J=\frac{1}{2}$, the TDVP equations of motion become
\begin{equation}
    \dot{\theta}_i =2 \Omega_i \sin\phi_i \cos\frac{\theta_{i+1}}{2} + \Omega_{i-1}\frac{\eta_{i-1}  \sin\frac{\theta_i}{2} \sin\phi_{i-1} \sin\theta_{i-1}}{ \eta_i}
\end{equation}
and
\begin{align}
    \dot{\phi}_i & =  2\Omega_i \cos\frac{\theta_{i+1}}{2} \cos\phi_i \cot\theta_i + 2 \Delta_i  - \Omega_{i-1}\frac{\cos\phi_{i-1} \sec\frac{\theta_{i}}{2} \sin\theta_{i-1} \eta_{i-1} }{2\eta_i}  \nonumber\\
    & - \Omega_i \frac{\cos\phi_i \sin\theta_i \sin \frac{\theta_{i+1}}{2} \eta_i \tan\frac{\theta_{i+1}}{2} }{2\eta_{i+1}} - \Omega_{i+1}\cos\frac{\theta_{i+2}}{2} \cos\phi_{i+1} \tan\frac{\theta_{i+1}}{2}
\end{align}
These equations provide a compact and efficient description of the TDVP dynamics for spin-$\frac{1}{2}$ system.

\section{Calculation of quantum leakage}\label{sec:qleak}
Quantum leakage quantifies the instantaneous error rate in variational quantum dynamics. The leakage rate, $\Gamma^2$, is expressed as 
\begin{align}
    \Gamma^2 &= \frac{1}{L}\Big( \bra{\Psi}H^2\ket{\Psi}_c - 2\sum_i \dot{\mu}_i \text{Im} \bra{\partial_{\mu_i}\Psi}H\ket{\Psi}_c + \sum_{ij} \dot{\mu}_i \text{Re} \inner{\partial_{\mu_i}\Psi}{\partial_{\mu_j}\Psi}_c \dot{\mu}_j \nonumber\\
    & = \bra{\Psi}H^2\ket{\Psi}_c - 2\sum_i \dot{\theta}_i \text{Im} \bra{\partial_{\theta_i}\Psi}H\ket{\Psi}_c- 2\sum_i \dot{\phi}_i \text{Im} \bra{\partial_{\phi_i}\Psi}H\ket{\Psi}_c\nonumber\\
    &+ \sum_{ij} \dot{\theta}_i \text{Re} \inner{\partial_{\theta_i}\Psi}{\partial_{\theta_j}\Psi}_c \dot{\theta}_j +  \sum_{ij} \dot{\phi}_i \text{Re} \inner{\partial_{\phi_i}\Psi}{\partial_{\phi_j}\Psi}_c \dot{\phi}_j  \Big),
\end{align}
where the subscript ``c'' denotes the connected part of the expectation value. The parameters $\mu_i$ represent the variational parameters.

Before addressing quantum leakage, let us shift focus to discuss the residual terms, which are crucial for the simplification of quantum leakage. In previous section, we define residual terms of $\text{Re} \bra{\partial_{\mu_i}\Psi}H\ket{\Psi}_c $ that exclude contributions involving  $(\text{Im} G^{c}_{\bm{\theta},\bm{\phi}})$ in Eq.~(\ref{eq:resi-phiH}) and (\ref{eq:resi-thetaH}), i.e.
\begin{align}
    R_{\phi_j} = \frac{L}{K}\Big( -\eta_j \frac{\tan\phi_j}{2} \Omega_j\cos\phi_j (\sin\theta_j +2x_j^2(-1+x_{j+1})\tan\frac{\theta_j}{2}) \Big)
\end{align}
\begin{align}
    R_{\theta_j}  &= \frac{L}{K}\Big( -J \Omega_{j-1} \tan\frac{\theta_j}{2} x_{j-1}^2 \tan\frac{\theta_{j-1}}{2}\cos\phi_{j-1} x_j \eta_{j-1} \nonumber\\
    & - \eta_j \Omega_j \cos\phi_j \Big( \frac{1}{2}+  x_j^2 (-1+x_{j+1}) (1 -2J + \frac{4J-1}{2\cos^2\frac{\theta_j}{2}})  \Big) \Big)
\end{align}
Similarly, we define the residual term of $\text{Im} \bra{\partial_{\mu_i}\Psi}H\ket{\Psi}_c $ that excludes the contribution involving $G^{c}_{\phi_i,\phi_j}$
\begin{align}
I_{\theta_j}  \equiv  \frac{L}{K}\Big( J \Omega_{j-1} \tan\frac{\theta_{j}}{2} \tan\frac{\theta_{j-1}}{2}\sin\phi_{j-1}  x^2_{j-1}x_j \eta_{j-1} 
+\eta_j \frac{\tan\phi_j}{2\sin\theta_j}\Omega_j\tilde{h}_{s_j^x} \Big),
\end{align}
\begin{equation}
    I_{\phi_j} =
- \frac{L}{K}\Big( \frac{\eta_j}{2} \Omega_j \cos\phi_j \Big( \sin\theta_j + (\cos\theta_j +2J(1-\cos\theta_j) ) 2x_j^2 (-1+x_{j+1})\tan\frac{\theta_j}{2}\Big) \Big),
\end{equation}
where the shorthand notation $\tilde{h}_{s^{x}_i } = \cos\phi_i (\sin\theta_i +2x_i^2(-1+x_{i+1})\tan\frac{\theta_i}{2})$ comes from Eq.~(\ref{eq:h-def}). These residual terms satisfy important relations
\begin{equation}
     R_{\theta_{j}} = \frac{L}{K}\Big( -J \Omega_{j-1} \tan\frac{\theta_j}{2} x_{j-1}^2 \tan\frac{\theta_{j-1}}{2}\cos\phi_{j-1} x_j \eta_{j-1} \Big) + \frac{ I_{\phi_j} }{\sin\theta_j},
\end{equation}
and
\begin{equation}
    I_{\theta_j}  =  \frac{L}{K}\Big( J \Omega_{j-1} \tan\frac{\theta_{j}}{2} \tan\frac{\theta_{j-1}}{2}\sin\phi_{j-1}  x^2_{j-1}x_j \eta_{j-1} \Big)
-\frac{R_{\phi_j}}{\sin\theta_j}.
\end{equation}

We now calculate quantum leakage. With the Gram matrix, variational dynamics and $\bra{\partial_{\mu_i}\Psi}H\ket{\Psi}_c$ already determined, the quantum leakage can be resolved once the variance of the Hamiltonian is calculated. The variance $\bra{\Psi}H^2\ket{\Psi}_c$ is bilinear in the spin operators $s^x$ and $s^x$. Then we can decompose the variance into ZZ, ZX, XZ and XX terms. The first three terms involves the correlation between onsite operators, which are for $|j-i|>1$
\begin{align}
    &(l_i|T_{q_i}\cdot T_{[i+1,j-1]}\cdot T_{q_j}|r_{j+1})_c \nonumber\\
    & = (l_i|T_{q_i}\cdot T_{[i+1,j-1]}\cdot T_{q_j}|r_{j+1}) -(l_i|T_{q_i}|r_{i+1})\cdot  (l_j|T_{q_j}|r_{j+1}) \nonumber\\
    & = -\tilde{h}_{q_i} \tilde{h}_{q_j} \eta_i\eta_{i+1} \beta_{[i+1,j-1]},
    \label{eq:corre-qiqj}
\end{align}
and for $j=i+1$
\begin{align}
    &(l_i|T_{q_i}\cdot T_{q_{i+1}}|r_{i+2})_c \nonumber\\
    & = (l_i|T_{q_i}\cdot T_{q_{i+1}}|r_{i+2}) -(l_i|T_{q_i}|r_{i+1})\cdot  (l_{i+1}|T_{q_{i+1}}|r_{i+2}) \nonumber\\
    & = \eta_i x_i x_{i+2} \Big(\bra{\varOmega}_iQ_iq_{i}\ket{0}_i\bra{0}_{i+1}q_{i+1}Q_{i+1}\ket{\varOmega}_{i+1} \nonumber\\
    &+\bra{0}_{i}q_{i}Q_{i}\ket{\varOmega}_{i}\bra{\varOmega}_{i+1}Q_{i+1}q_{i+1}\ket{0}_{i+1} \Big) -\tilde{h}_{q_i} \tilde{h}_{q_{i+1}} \eta_i\eta_{i+1}.
\end{align}
The self-term $j=i$ contribution is essentially the expectation of single onsite operators, which is presented in Eq.~(\ref{eq:avq}). For the XX term, special case is required. When the distance $|j-i|$ between two spin operators $s^x_i$ and $s^x_j$ is larger than one, the correlation reduces to Eq.~(\ref{eq:corre-qiqj}). The nontrivial case occurs when $|j-i|\leq 1$. From $\bra{\Psi} \mathcal{P} s_i^x \mathcal{P} q_{i+1}  \ket{\Psi}  = \bra{\Psi} P_{i-1,i}P_{i,i+1} s^x_i  P_{i-1,i}P_{i,i+1} q_{i+1}   \ket{\Psi} = \bra{\Psi} P_{i-1} s^x_i  P_{i+1}  q_{i+1} \ket{\Psi}$, we can omit $\mathcal{P}$ by using the dressed operator $P_{i-1}s_i^x P_{i+1}$. The XX term involves for $|j-i|=1$
\begin{align}
    &(l_{i-1}|T_{P_{i-1}}\cdot T_{s_i^x}\cdot T_{P_{i+1}q_{i+1}}|r_{i+2})_c \nonumber\\
    & = (l_{i-1}|T_{P_{i-1}}\cdot T_{s_i^x}\cdot T_{P_{i+1}q_{i+1}}|r_{i+2}) -(l_i|T_{s_i^x}|r_{i+1})\cdot  (l_{i+1}|T_{q_{i+1}}|r_{i+2}) \nonumber\\
    & = -\tilde{h}_{s_i^x} \tilde{h}_{q_{i+1}} \eta_i \eta_{i+1} + \eta_i x_i x_{i+2}\bra{\varOmega}_iQ_is_{i}^x\ket{0}_i\bra{0}_{i+1}q_{i+1}Q_{i+1}\ket{\varOmega}_{i+1},  
\end{align}
\begin{align}
    &(l_{i}|T_{P_{i-1}}\cdot T_{s_i^x}\cdot T_{P_{i+1}}|r_{i+2})_c \nonumber\\
    & = (l_{i-1}|T_{P_{i-1}q_{i-1}}\cdot T_{s_i^x}\cdot T_{P_{i+1}}|r_{i+2}) - (l_i|T_{s_i^x}|r_{i+1})\cdot  (l_{i-1}|T_{q_{i-1}}|r_{i}) \nonumber\\
    & = -\tilde{h}_{s_i^x} \tilde{h}_{q_{i-1}} \eta_i \eta_{i-1} + \eta_{i-1} x_{i-1} x_{i+1} \bra{\varOmega}_iQ_is_{i}^x\ket{0}_i\bra{0}_{i-1}q_{i-1}Q_{i-1}\ket{\varOmega}_{i-1},  
\end{align}
and for $j=i$ case
\begin{align}
    &(l_{i-1}|T_{P_{i-1}}\cdot T_{(s_i^x)^2}\cdot T_{P_{i+1}}|r_{i+2})_c \nonumber\\
    & = (l_{i-1}|T_{P_{i-1}}\cdot T_{(s_i^x)^2}\cdot T_{P_{i+1}}|r_{i+2}) - (l_i|T_{s_i^x}|r_{i+1})^2 \nonumber\\
    & = \eta_i (\tilde{h}_{(s_i^x)^2} +(1+x_i^2(-1+x_{i+1}^2))\bra{0}_i (s_i^x)^2 \ket{0}_i ) - (\eta_i \tilde{h}_{s_i^x})^2.  
\end{align}
Collecting these equations, and substituting results from Tables \ref{tab:one-site op} and \ref{tab:2-site op} into ZZ, ZX, XZ and XX terms, we can solve the energy variance. Explicitly, the ZZ term is
\begin{align}
    \text{Var}_{ZZ} &= -\frac{L}{K}\sum_{i=1}^{K} \sum_{j=i+1}^{i+K}\Delta_i \Delta_j  (1-\cos\theta_i)(1-\cos\theta_j)\frac{\eta_i\eta_{i+1}\beta_{[i+1,j-1]}}{1-\beta_{[1,K]}} \nonumber\\
    &-\frac{L}{K}\sum_{i=1}^{K}\sum_{j=i-K}^{i-1} \Delta_i \Delta_j  (1-\cos\theta_i)(1-\cos\theta_j) \frac{\eta_j\eta_{j+1}\beta_{[j+1,i-1]}}{1-\beta_{[1,K]}} \nonumber\\
    &+ \frac{L}{K}\sum_{i=1}^{K} \Delta_i^2 \eta_i \frac{1+\cos\theta_i-2J(-1+\eta_i)(1-\cos\theta_i)}{J}\sin^2\frac{\theta_i}{2}.
\end{align}
One can see that
\begin{align}
    (\text{Var}_{ZZ})_{ij} = \frac{\Delta_i \Delta_j}{J^2} G^{c}_{\phi_i,\phi_j}.
\end{align}
The ZX term is
\begin{align}
    \text{Var}_{ZX} &= -\frac{L}{K}\sum_{i=1}^{K} \sum_{j=i+1}^{i+K}\eta_i\eta_{i+1} \Delta_i \Omega_{j} (1-\cos\theta_i)\Big(\sin\theta_j \cos\phi_j \nonumber\\
    &+2(-1+x_{j+1})x_j^2 \tan\frac{\theta_j}{2}\cos\phi_j\Big)\frac{\beta_{[i+1,j-1]}}{1-\beta_{[1,K]}}\nonumber\\
    & - \frac{L}{K}\sum_{i=1}^{K} \sum_{j=i-K}^{i-1}\eta_j\eta_{j+1} \Delta_i \Omega_{j} (1-\cos\theta_i)\Big(\sin\theta_j \cos\phi_j \nonumber\\
    &+2(-1+x_{j+1})x_j^2 \tan\frac{\theta_j}{2}\cos\phi_j\Big)\frac{\beta_{[j+1,i-1]}}{1-\beta_{[1,K]}} \nonumber\\
    &+\frac{L}{K}\sum_{i=1}^{K} \Delta_i \Omega_{i} \frac{\eta_i }{2J}\Big( \sin\theta_i\cos\phi_i (\cos\theta_i+2J(1-\eta_i)(1-\cos\theta_i))
    \nonumber\\
    &+\ii \sin\theta_i \sin\phi_i +2\ii \sin\phi_i x_i^2(-1+x_{i+1})\tan\frac{\theta_i}{2} \nonumber\\
    &+ 2(1-2J\eta_i+2J\eta_i\cos\theta_i)\cos\phi_i x_i^2(-1+x_{i+1})\tan\frac{\theta_i}{2}\Big).
\end{align}
Notice for $i\neq j$
\begin{align}
    (\text{Var}_{ZX})_{ij} = \frac{\Delta_i \Omega_j}{J^2} G^{c}_{\phi_i,\phi_j}\frac{\sin\theta_j \cos\phi_j+2(-1+x_{j+1})x_j^2 \tan\frac{\theta_j}{2}\cos\phi_j}{1-\cos\theta_j}.
\end{align}
The imaginary parts of ZX and XZ terms are canceled, since $\text{Var}_{XZ} = \text{Var}_{ZX}^{*}$
\begin{align}
    &(\text{Var}_{ZX})+(\text{Var}_{XZ})\nonumber\\
    & = 2\sum_{i=1}^{K} \sum_{j=i}^{i+K-1} \frac{\Delta_i }{J^2} G^{c}_{\phi_i,\phi_j}\frac{\Omega_j \tilde{h}_{ s^{x}_j} }{1-\cos\theta_j} \nonumber\\
    & -2\frac{L}{K}\sum_{i=1}^{K} \frac{\eta_i}{2J}\Delta_i\Omega_i \cos\phi_i \Big( \sin\theta_i  + (\cos\theta_i +2J(1-\cos\theta_i))2x_i^2 (-1+x_{i+1}) \tan\frac{\theta_i}{2} \Big) \nonumber\\
    & =  2\sum_{i=1}^{K} \sum_{j=i}^{i+K-1} \frac{\Delta_i }{J^2} G^{c}_{\phi_i,\phi_j}\frac{\Omega_j \tilde{h}_{ s^{x}_j}}{1-\cos\theta_j} + 2\sum_{i=1}^{K} \frac{\Delta_i}{J}I_{\phi_i}. 
\end{align}
The last equality can be derived from Eq.~(\ref{P8eq:imdphiH}). 
The XX term in $\bra{\Psi}H^2\ket{\Psi}_c$ is
\begin{align}
    \text{Var}_{XX} & = \sum_{i=1}^{K} \sum_{j=i}^{i+K-1} \frac{G^{c}_{\phi_i,\phi_j}}{J^2}  \frac{\Omega_i \tilde{h}_{ s^{x}_i} }{1-\cos\theta_i} \frac{\Omega_j \tilde{h}_{ s^{x}_j}}{1-\cos\theta_j} \nonumber\\
    & + 2\frac{L}{K}\sum_{i=1}^{K} \Omega_i \Omega_{i+1} \eta_i x_i^2 x_{i+1} x_{i+2} \tan\frac{\theta_i}{2} \tan\frac{\theta_{i+1}}{2} \cos(\phi_{i+1}-\phi_i)  \nonumber\\
    &+ \frac{L}{K}\sum_{i=1}^{K} \frac{\eta_i \Omega_i^2}{2J} \Big( 1+ (-1+2J)\cos^2\phi_i\sin^2\theta_i \nonumber\\
    & + \frac{(-1-2J+(-1+2J)\cos\theta_i)( \tilde{h}_{ s^{x}_i})^2}{1-\cos\theta_i} \nonumber\\
    & + x_i^2 (-1+x_{i+1})(1+x_{i+1}+2(-1+2J)\cos2\phi_i \tan^2\frac{\theta_i}{2})\Big). 
\end{align}
Finally we have the variance of the Hamiltonian $\bra{\Psi}H^2\ket{\Psi}_c = \text{Var}_{XX}+\text{Var}_{ZX}+\text{Var}_{XZ}+\text{Var}_{ZZ}$, which is
\begin{align}
    &\bra{\Psi}H^2\ket{\Psi}_c \nonumber\\
    &=\sum_{i=1}^{K} \sum_{j=i}^{i+K-1} \frac{G^{c}_{\phi_i,\phi_j}}{J^2}  \frac{ \Omega_i\tilde{h}_{ s^{x}_i}+\Delta_i\tilde{h}_{ s^{z}_i} }{1-\cos\theta_i} \frac{ \Omega_j\tilde{h}_{ s^{x}_j}+\Delta_j\tilde{h}_{ s^{z}_j} }{1-\cos\theta_j} + 2\sum_{i=1}^{K} \frac{\Delta_i}{J}I_{\phi_i} \nonumber\\
    & + 2\frac{L}{K}\sum_{i=1}^{K} \Omega_i \Omega_{i+1} \eta_i x_i^2 x_{i+1} x_{i+2} \tan\frac{\theta_i}{2} \tan\frac{\theta_{i+1}}{2} \cos(\phi_{i+1}-\phi_i)  \nonumber\\
    & +\frac{L}{K}\sum_{i=1}^{K} \frac{\eta_i \Omega_i^2}{2J} \Big( 1+ (-1+2J)\cos^2\phi_i\sin^2\theta_i \nonumber\\
    & + \frac{(-1-2J+(-1+2J)\cos\theta_i)( \tilde{h}_{ s^{x}_i})^2}{1-\cos\theta_i}  \nonumber\\
    & + x_i^2 (-1+x_{i+1})(1+x_{i+1}+2(-1+2J)\cos2\phi_i \tan^2\frac{\theta_i}{2})\Big), 
\end{align}
where $\Omega_i\tilde{h}_{ s^{x}_i}+\Delta_i\tilde{h}_{ s^{z}_i} = \Omega_i \cos\phi_i (\sin\theta_i +2x_i^2(-1+x_{i+1})\tan\frac{\theta_i}{2}) + \Delta_i (1-\cos\theta_i)$ comes from Eq.~(\ref{eq:h-def}).

Next, we solve other terms in the quantum leakage, which involve the TDVP equations of motion Eq.~(\ref{eq:eom-theta}) and (\ref{eq:eom-phi}).  There are two terms involving $\text{Im} \bra{\partial_{\mu_i}\Psi}H\ket{\Psi}_c $
\begin{align}
\sum_{i=1}^{K} \dot{\theta}_i \text{Im} \bra{\partial_{\theta_i}\Psi}H\ket{\Psi}_c 
= \sum_{i}\sum_{j\leq i}(\text{Im} G^{c}_{\bm{\theta},\bm{\phi}})_{ji}^{-1} \cdot R_{\phi_j} \cdot I_{\theta_i},
\end{align}
\begin{align}
    &\sum_{i=1}^{K} \dot{\phi}_i \text{Im} \bra{\partial_{\phi_i}\Psi}H\ket{\Psi}_c \nonumber\\
    & = \sum_i \bigg( \sum_{j=i}^{i+K-1} \frac{G^{c}_{\phi_i,\phi_j}}{J^2}  \frac{ \Omega_i\tilde{h}_{ s^{x}_i}+\Delta_i\tilde{h}_{ s^{z}_i} }{1-\cos\theta_i} \frac{ \Omega_j\tilde{h}_{ s^{x}_j}+\Delta_j\tilde{h}_{ s^{z}_j} }{1-\cos\theta_j} \nonumber\\
    & -\sum_{j,l} (\text{Im} G^{c}_{\bm{\theta},\bm{\phi}})_{ij}^{-1} \cdot R_{\theta_j}\cdot \frac{G^{c}_{\phi_i,\phi_l} (\Omega_l\tilde{h}_{ s^{x}_l}+\Delta_l\tilde{h}_{ s^{z}_l})  }{J(1-\cos\theta_l)} + \dot{\phi}_i \cdot I_{\phi_i} \bigg).
\end{align}
There are two terms involving the diagonal part of the Gram matrix
\begin{align}
    \sum_{i,j} \dot{\theta}_i  G^{c}_{\theta_i,\theta_j} \dot{\theta}_j = \frac{L}{K}\sum_{i=1}^{k} \frac{\eta_i J}{2} \Big(  \sum_{j\leq i} (\text{Im} G^{c}_{\bm{\theta},\bm{\phi}})_{ji}^{-1} R_{\phi_j}\Big)^2,
\end{align}
\begin{align}
    &\sum_{i,j} \dot{\phi}_i  G^{c}_{\phi_i,\phi_j} \dot{\phi}_j \nonumber\\
    & = \sum_{i=1}^{K} \sum_{j=i}^{i+K-1} \Big( \frac{G^{c}_{\phi_i,\phi_j}}{J^2}  \frac{  \Omega_i\tilde{h}_{ s^{x}_i}+\Delta_i\tilde{h}_{ s^{z}_i}  }{1-\cos\theta_i} \frac{  \Omega_j\tilde{h}_{ s^{x}_j}+\Delta_j\tilde{h}_{ s^{z}_j}  }{1-\cos\theta_j} \nonumber\\
    & -2 \frac{  \Omega_i\tilde{h}_{ s^{x}_i}+\Delta_i\tilde{h}_{ s^{z}_i}  }{J(1-\cos\theta_i)} G^{c}_{\phi_i,\phi_j} \sum_l (\text{Im} G^{c}_{\bm{\theta},\bm{\phi}})_{jl}^{-1} \cdot  R_{\theta_l}  \nonumber\\
    & + \sum_k (\text{Im} G^{c}_{\bm{\theta},\bm{\phi}})_{ik}^{-1} \cdot  R_{\theta_k} G^{c}_{\phi_i,\phi_j} \sum_l (\text{Im} G^{c}_{\bm{\theta},\bm{\phi}})_{jl}^{-1} \cdot  R_{\theta_l} \Big). 
\end{align}
Combining these together, we obtain the quantum leakage
{\allowdisplaybreaks
\begin{align}
    \Gamma^2 &= 2\frac{1}{K}\sum_{i=1}^{K} \Omega_i \Omega_{i+1} \eta_i x_i^2 x_{i+1} x_{i+2} \tan\frac{\theta_i}{2} \tan\frac{\theta_{i+1}}{2} \cos(\phi_{i+1}-\phi_i)  \nonumber\\
    & +\frac{1}{K}\sum_{i=1}^{K} \frac{\eta_i \Omega_i^2}{2J} \Big( 1+ (-1+2J)\cos^2\phi_i\sin^2\theta_i \nonumber\\
    & + \frac{(-1-2J+(-1+2J)\cos\theta_i)( \tilde{h}_{ s^{x}_i})^2}{1-\cos\theta_i}  \nonumber\\
    & + x_i^2 (-1+x_{i+1})(1+x_{i+1}+2(-1+2J)\cos2\phi_i \tan^2\frac{\theta_i}{2})\Big) \nonumber\\
    & -2 \frac{1}{L} \sum_{i}\sum_{j\leq i}(\text{Im} G^{c}_{\bm{\theta},\bm{\phi}})_{ji}^{-1} \cdot R_{\phi_j} \cdot I_{\theta_i}  \nonumber\\
    & - 2 \sum_i\Big(  \frac{\Omega_i\tilde{h}_{ s^{x}_i}}{J(1-\cos\theta_i)}  \frac{1}{L}I_{\phi_i}  - \sum_j (\text{Im} G^{c}_{\bm{\theta},\bm{\phi}})_{ij}^{-1} \cdot R_{\theta_j}  \frac{1}{L} I_{\phi_i} \Big)\nonumber\\
    & + \frac{1}{K}\sum_{i=1}^{k} \frac{\eta_i J}{2} \Big(  \sum_j(\text{Im} G^{c}_{\bm{\theta},\bm{\phi}})_{ji}^{-1} R_{\phi_j}\Big)^2  \nonumber\\
    & + \frac{1}{L}\sum_{ij}\sum_k (\text{Im} G^{c}_{\bm{\theta},\bm{\phi}})_{ik}^{-1} \cdot  R_{\theta_k} \cdot G^{c}_{\phi_i,\phi_j} \sum_l (\text{Im} G^{c}_{\bm{\theta},\bm{\phi}})_{jl}^{-1} \cdot  R_{\theta_l},
    \label{eq:qleak-rough}
\end{align}
}
where we have use the fact that $G^{c}_{\phi_i,\phi_j}$ is symmetric matrix.
It is interesting to note that quantum leakage is independent of detuning $\Delta$, since
\begin{equation}
    \frac{\partial \Gamma^2}{\partial \Delta_i} =0.
\end{equation}

The last term in Eq.~(\ref{eq:qleak-rough}) can be further simplified. Note that the diagonal element $G^{c}_{\phi_i,\phi_i} $ satisfies
\begin{equation}
     G^{c}_{\phi_i,\phi_i} -(-\frac{\eta_{i+1}\sin\theta_i}{x_i^2}(\text{Im} G^{c}_{\bm{\theta},\bm{\phi}})_{ii}) = \frac{L}{K} \eta_i 2J \sin^2\frac{\theta_i}{2}(1-\eta_i) (2J\sin^2\frac{\theta_i}{2}+\cos^2\frac{\theta_i}{2} - \frac{\cos^2\frac{\theta_i}{2}}{x_i^2}),
\end{equation}
which can be generalized for $i\leq j$
\begin{equation}
    \frac{\eta_{i+1}\sin\theta_i}{x_i^2}(\text{Im} G^{c}_{\bm{\theta},\bm{\phi}})_{ij} + G^{c}_{\phi_i,\phi_j} = \frac{L}{K} \delta_{ij}\eta_i 2J \sin^2\frac{\theta_i}{2} (1-\eta_i)\frac{\cos^2\frac{\theta_i}{2}}{x_i^2} \tilde{c}_i.
\end{equation}
Here $\tilde{c}_i = -1+ (2J \tan^2\frac{\theta_i}{2}+1)x_i^2$, see Eq.~(\ref{eq:parameter-in-invG}).
Therefore we have the decomposition
\begin{align}
     G^{c}_{\phi_i,\phi_j} &= -\frac{\eta_{i+1}\sin\theta_i}{x_i^2}(\text{Im} G^{c}_{\bm{\theta},\bm{\phi}})_{ij} - \frac{\eta_{j+1}\sin\theta_j}{x_j^2}(\text{Im} G^{c}_{\bm{\theta},\bm{\phi}})_{ji} \nonumber\\
     & + \delta_{ij}\Big( \frac{\eta_{i+1}\sin\theta_i}{x_i^2}(\text{Im} G^{c}_{\bm{\theta},\bm{\phi}})_{ii} + \frac{L}{K} \eta_i 2J \sin^2\frac{\theta_i}{2} (1-\eta_i)\frac{\cos^2\frac{\theta_i}{2}}{x_i^2} \tilde{c}_i\Big) \nonumber\\
     & = -\frac{\eta_{i+1}\sin\theta_i}{x_i^2}(\text{Im} G^{c}_{\bm{\theta},\bm{\phi}})_{ij} - \frac{\eta_{j+1}\sin\theta_j}{x_j^2}(\text{Im} G^{c}_{\bm{\theta},\bm{\phi}})_{ji} \nonumber\\
     &+ \frac{L}{K}\delta_{ij} \frac{\eta_i J \sin^2\theta_i}{2x_i^2} \big( (1-\eta_i) \tilde{c}_i -\eta_{i+1} \big).
\end{align}
Substitute into the last term in Eq.~(\ref{eq:qleak-rough}), yielding
\begin{align}
    &\sum_{ij}\sum_k (\text{Im} G^{c}_{\bm{\theta},\bm{\phi}})_{ik}^{-1} \cdot  R_{\theta_k} \cdot G^{c}_{\phi_i,\phi_j} \sum_l (\text{Im} G^{c}_{\bm{\theta},\bm{\phi}})_{jl}^{-1} \cdot  R_{\theta_l}\nonumber\\
    & = 2\sum_{ik} (\text{Im} G^{c}_{\bm{\theta},\bm{\phi}})_{ik}^{-1} \cdot  R_{\theta_k} (-\frac{\eta_{i+1}\sin\theta_i}{x_i^2})  R_{\theta_i} \nonumber\\
    & + \frac{L}{K} \sum_i \Big( \frac{\eta_i J \sin^2\theta_i}{2x_i^2} \big( (1-\eta_i) \tilde{c}_i -\eta_{i+1} \big) \Big) \cdot \Big( \sum_{j} (\text{Im} G^{c}_{\bm{\theta},\bm{\phi}})_{ij}^{-1} \cdot  R_{\theta_j} \Big)^2.
\end{align}
Therefore, quantum leakage can be rewritten in terms of $ \sum_{j} (\text{Im} G^{c}_{\bm{\theta},\bm{\phi}})_{ij}^{-1} \cdot  R_{\theta_j}$ and $\sum_j(\text{Im} G^{c}_{\bm{\theta},\bm{\phi}})_{ji}^{-1} R_{\phi_j}$
{\allowdisplaybreaks
\begin{align}
    \Gamma^2 &= 2\frac{1}{K}\sum_{i=1}^{K} \Omega_i \Omega_{i+1} \eta_i x_i^2 x_{i+1} x_{i+2} \tan\frac{\theta_i}{2} \tan\frac{\theta_{i+1}}{2} \cos(\phi_{i+1}-\phi_i)  \nonumber\\
    & +\frac{1}{K}\sum_{i=1}^{K} \frac{\eta_i \Omega_i^2}{2J} \Big( 1+ (-1+2J)\cos^2\phi_i\sin^2\theta_i \nonumber\\
    & +\frac{(-1-2J+(-1+2J)\cos\theta_i)( \tilde{h}_{ s^{x}_i})^2}{1-\cos\theta_i} \nonumber\\
    & + x_i^2 (-1+x_{i+1})(1+x_{i+1}+2(-1+2J)\cos2\phi_i \tan^2\frac{\theta_i}{2})\Big) \nonumber\\
    & -2 \frac{1}{L} \sum_{i}\sum_{j\leq i}(\text{Im} G^{c}_{\bm{\theta},\bm{\phi}})_{ji}^{-1} \cdot R_{\phi_j} \cdot I_{\theta_i}  \nonumber\\
    & - 2 \sum_i\Big(  \frac{ \Omega_i\tilde{h}_{ s^{x}_i} }{J(1-\cos\theta_i)}  \frac{1}{L}I_{\phi_i}  - \sum_j (\text{Im} G^{c}_{\bm{\theta},\bm{\phi}})_{ij}^{-1} \cdot R_{\theta_j}  \frac{1}{L} I_{\phi_i} \Big)\nonumber\\
    & + \frac{1}{K}\sum_{i=1}^{k} \frac{\eta_i J}{2} \Big(  \sum_j(\text{Im} G^{c}_{\bm{\theta},\bm{\phi}})_{ji}^{-1} R_{\phi_j}\Big)^2  \nonumber\\
    & + 2\frac{1}{L} \sum_{ik} (\text{Im} G^{c}_{\bm{\theta},\bm{\phi}})_{ik}^{-1} \cdot  R_{\theta_k} (-\frac{\eta_{i+1}\sin\theta_i}{x_i^2})  R_{\theta_i} \nonumber\\
    & + \frac{1}{K} \sum_i \Big( \frac{\eta_i J \sin^2\theta_i}{2x_i^2} \big( (1-\eta_i) \tilde{c}_i -\eta_{i+1} \big) \Big) \cdot \Big( \sum_{j} (\text{Im} G^{c}_{\bm{\theta},\bm{\phi}})_{ij}^{-1} \cdot  R_{\theta_j} \Big)^2.
\end{align}
}
When $J=\frac{1}{2}$, the expression is surprisingly simple
\begin{equation}
    \Gamma^2 = \frac{1}{K}\sum_{i=1}^{K} \Omega_i^2 \sin^2\frac{\theta_i}{2} \sin^2\frac{\theta_{i+1}}{2} \frac{\eta_i (1-\eta_i)}{\eta_{i+1}}.
\end{equation}



\providecommand{\noopsort}[1]{}\providecommand{\singleletter}[1]{#1}%

\end{document}